\documentstyle[12pt,epsf,rotate]{article}
\if@twoside \oddsidemargin 21pt \evensidemargin 59pt \marginparwidth 85pt
\else \oddsidemargin 0pt \evensidemargin 0pt
\fi
\renewcommand{\vec}[1]{\mbox{\boldmath $#1$}}
\newcommand{\weg}[1]{#1}
\newcounter{saveeqn}                  
\newcommand{\alpheqn}[1]{\refstepcounter{equation}\label{#1}%
\setcounter{saveeqn}{\value{equation}}%
\setcounter{equation}{0}%
\renewcommand{\theequation}
{\mbox{\arabic{saveeqn}\alph{equation}}}}
\newcommand{\reseteqn}{\setcounter{equation}{\value{saveeqn}}%
\renewcommand{\theequation}{\arabic{equation}}}
\newcounter{savefig}

\begin{document}
\title{{\sc Boltzmann}-like and {\sc Boltzmann-Fokker-Planck} Equations
as a Foundation of Behavioral Models}

\author{Dirk Helbing} 
\maketitle
\vfill
\begin{abstract}
\mbox{ }\\[-0.5cm]
It is shown, that the {\sc Boltzmann}-like equations allow the formulation
of a very general model for behavioral changes. This model takes into
account spontaneous (or externally induced) behavioral changes and
behavioral changes by pair interactions. As most important social pair
interactions imitative and avoidance processes are distinguished. The
resulting model turns out to include as special cases
many theoretical concepts of the social sciences.\\[2mm] 
A {\sc Kramers-Moyal} expansion of the {\sc Boltzmann}-like equations leads
to the {\sc Boltzmann-Fokker-Planck} equations, which allows the introduction
of ``social forces'' and ``social fields''. A social field reflects the
influence of the public opinion, social norms and trends 
on behaviorial changes. It is not only given by
external factors (the environment)
but also by the interactions of the individuals. 
Variations of the individual behavior are taken into 
account by diffusion coefficients.
\end{abstract}
\clearpage
\setcounter{page}{1}
\section{Introduction} \label{s1}

The methods of statistical physics have shown to be very fruitful 
in physics, but in the last decades they also have become increasingly
important in interdisciplinary research. For example, the {\em master
equation} has found many applications in thermodynamics 
\cite{Zwan}, chemical kinetics \cite{Opp}, 
laser theory \cite{Hak} and biology \cite{Arn}. 
Moreover, {\sc Weid\-lich} and
{\sc Haag} have successfully introduced it for the description of 
social processes \cite{WeHa83,We91} like opinion formation \cite{Weid1}, 
migration \cite{Weid2}, agglomeration \cite{WeHa87} and
settlement processes \cite{WeMu90}. 
\par
Another kind of wide-spread equations are the {\sc Boltzmann} equations,
which have been developed for the description of the kinetics
of gases \cite{Bo64} and of chemical reactions 
\cite{Br82}. However, {\sc Boltzmann}-{\em like equations} \cite{Helb}
play also an important role for quantitative models in the
social sciences: It turns out (cf. sect. \ref{s2.2}) that the 
{\em logistic equation} for the description of limited
growth processes \cite{Pea24,Ve45}, 
the socalled {\em gravity model} for spatial exchange processes
\cite{Zi46}, and the {\em game dynamical equations} modelling 
competition and cooperation processes \cite{HoSi88,Mue90} are special cases
of {\sc Boltzmann}-like equations.
Moreover, {\sc Boltzmann}-like models have recently been 
suggested for avoidance processes of 
pedestrians \cite{Diss,Compl} and for attitude formation by direct pair
interactions of individuals occuring in discussions \cite{Diss,He92}. 
\par
In this paper we shall show that {\sc Boltzmann}-like equations and
{\sc Boltzmann-Fokker-Planck} equations \cite{Helb} are suited as a
foundation of quantitative behavioral models. For this purpose, we shall
proceed in the following way:
In section \ref{s2} the {\sc Boltzmann}-like equations will be introduced
and applied to the description of behavioral changes. The model includes
{\em spontaneous} (or {\em externally induced}) 
behavioral changes and behavioral changes by
{\em pair interactions} of individuals. 
These changes are described 
by {\em transition rates}. They reflect the results of mental
and psychical processes, which could be simulated with help
of {\sc Osgood} and {\sc Tannenbaum}'s 
{\em congruity principle} \cite{OsTa55}, {\sc Heider}'s 
{\em balance theory} \cite{He46} or {\sc Festinger}'s {\em dissonance
theory} \cite{Fe57}.
However, it is sufficient for our model to determine the transition rates
empirically (sect. \ref{s5}). The {\em ansatz} 
used for the transition rates distinguishes
{\em imitative} and {\em avoidance processes}, 
and assumes {\em utility maximization} of the individuals
(sect. \ref{s2.1}).
It is shown, that the resulting
{\sc Boltzmann}-like model for imitative processes
implies as special cases many generally accepted theoretical approaches 
in the social sciences (sect. \ref{s2.2}).
\par
In section \ref{s3} a consequent mathematical 
formulation related to an idea of {\sc Lewin} \cite{Le51}
is developed, according to which the behavior of individuals is
guided by a {\em social field}. This formulation is achieved by a
{\sc Kramers-Moyal} expansion of the {\sc Boltzmann}-like equations
leading to a kind of {\em diffusion equations}: 
the socalled {\sc Boltzmann-Fokker-Planck} {\em equations} \cite{Helb}. 
In these equations the most
probable behavioral change is given by a 
vectorial quantity that can be interpreted
as {\em social force} (sect. \ref{s3.1}). 
The social force results from external
influences (the environment) as well as from individual 
interactions. In special cases the social 
force is the gradient of
a potential. This potential reflects the public opinion, 
social norms and trends, and will be called the {\em social field}. 
By {\em diffusion coefficients} an individual variation of the behavior
(the ``freedom of will'') is taken into account. In section \ref{s4} 
representative cases are illustrated by computer simulations. 
\par
The {\sc Boltzmann-Fokker-Planck} modell for the behavior of individuals under
the influence of a social field shows some analogies with the physical model
for the behavior of electrons in an electric field (e.g. of an atomic
nucleus) \cite{Helb,Diss}
(cf. {\sc Hartree}'s {\em selfconsistent field ansatz} \cite{Ha28}). 
Especially, individuals and electrons influence the concrete
form of the {\em effective} social resp. electric field.
However, the behavior of electrons is governed by a {\em different}
equation: the {\sc Schr\"odinger} {\em equation}.
\par
In physics, the {\sc Boltzmann-Fokker-Planck} equations can be used
for the description of {\em diffusion} processes \cite{Mo71}.

\section{The {\sc Boltzmann}-like equations} \label{s2}

Let us consider a {\em system} consisting
of a great number $N\gg 1$ of {\em subsystems}.
These subsystems are in one {\em state} $\vec{x}$ of several possible
states combined in the set $\Omega$. 
\par
Due to {\em fluctuations} one cannot expect a deterministic theory
for the temporal change $d\vec{x}/dt$ of the state
$\vec{x}(t)$ to be realistic. However, one can construct a {\em stochastic}
model for the change of the {\em probability distribution} $P(\vec{x},t)$ of
states $\vec{x}(t)$ within the given system ($P(\vec{x},t) \ge 0$,
$\displaystyle \sum_{\weg{x}\in \Omega} P(\vec{x},t) = 1$). 
By introducing an index $a$ we may distinguish $A$ different {\em types}
$a$ of subsystems. If $N_a$ denotes the number of subsystems of type $a$,
we have $\displaystyle \sum_{a=1}^A N_a = N$, and the following relation holds:
\begin{equation}
 P(\vec{x},t) = \sum_{a=1}^A \frac{N_a}{N} P_a(\vec{x},t) \, .
\end{equation}
Our goal is now to find a suitable equation for the probability distribution
$P_a(\vec{x},t)$ of states for subsystems of type $a$
($P_a(\vec{x},t) \ge 0$, $\displaystyle \sum_{\weg{x}\in \Omega} 
P_a(\vec{x},t) = 1$). If we neglect
memory effects (cf. sect. \ref{s6.1}), the desired equation has the form
of a {\em master equation} \cite{Helb,Diss}:
\begin{equation}
 \frac{d}{dt} P_a(\vec{x},t) = \sum_{\weg{x}'\in \Omega \atop
 (\weg{x}' \ne \weg{x})} \Big[
 w^a(\vec{x}|\vec{x}';t) P_a(\vec{x}',t) 
 - w^a(\vec{x}'|\vec{x};t) P_a(\vec{x},t) \Big] \, .
\label{Boltz}
\end{equation}
$w^a(\vec{x}'|\vec{x};t)$ is the {\em effective transition rate}
from state $\vec{x}$ to $\vec{x}'$ and takes into account 
the fluctuations. Restricting the model to 
spontaneous (or externally induced)
transitions and transitions due to pair
interactions, we have \cite{Helb,Diss}:
\begin{equation}
 w^a(\vec{x}'|\vec{x};t) := w_a(\vec{x}'|\vec{x};t)
+ \sum_{b=1}^A \sum_{\weg{y}\in \Omega} \sum_{\weg{y}'\in \Omega} 
 N_b \, \widetilde{w}_{ab}
(\vec{x}',\vec{y}'|\vec{x},\vec{y};t) P_b(\vec{y},t) \, .
\label{effrate}
\end{equation}
$w_a(\vec{x}'|\vec{x};t)$ describes the rate of spontaneous 
(resp. externally induced) transitions from
$\vec{x}$ to $\vec{x}'$ for subsystems of type $a$.
$\widetilde{w}_{ab}(\vec{x}',\vec{y}'|\vec{x},\vec{y};t)$
is the transition rate for two subsystems of types $a$ and $b$
to change their states from $\vec{x}$ and $\vec{y}$ to
$\vec{x}'$ and $\vec{y}'$ due to pair interactions. 
\par
Inserting (\ref{effrate}) into (\ref{Boltz}), we now obtain the socalled
{\sc Boltzmann}-{\em like equations} \cite{Helb,Diss}\alpheqn{Boltzlike}
\begin{eqnarray}
 \frac{d}{dt} P_a(\vec{x},t) &=& \sum_{\weg{x}'\in \Omega} \Big[
 w_a(\vec{x}|\vec{x}';t) P_a(\vec{x}',t) 
 - w_a(\vec{x}'|\vec{x};t) P_a(\vec{x},t) \Big] \\
 &+& \sum_{b=1}^A \sum_{\weg{x}'\in \Omega} \sum_{\weg{y}\in \Omega} 
 \sum_{\weg{y}'\in \Omega} 
 w_{ab}(\vec{x},\vec{y}'|\vec{x}',\vec{y};t) P_b(\vec{y},t)P_a(\vec{x}',t)
 \nonumber \\
 &-& \sum_{b=1}^A \sum_{\weg{x}'\in \Omega} \sum_{\weg{y}\in \Omega} 
 \sum_{\weg{y}'\in \Omega} 
 w_{ab}(\vec{x}',\vec{y}'|\vec{x},\vec{y};t) P_b(\vec{y},t)P_a(\vec{x},t)
\end{eqnarray}\reseteqn
with
\begin{equation}
 w_{ab}(\vec{x}',\vec{y}'|\vec{x},\vec{y};t)
 := N_b \, \widetilde{w}_{ab}(\vec{x}',\vec{y}'|\vec{x},\vec{y};t) \, .
\end{equation}
Obviously, (\ref{Boltzlike}b) depends nonlinearly on the
probability distributions $P_a(\vec{x},t)$, which
is due to the interaction processes. 
\par
Neglecting spontaneous transitions (i.e., $w_a(\vec{x}'|\vec{x};t) \equiv 0$)
the {\sc Boltzmann}-like equations agree with the {\sc Boltzmann} equations,
that originally have been developed for the
description of the kinetics of gases \cite{Bo64}. A more detailed
discussion can be found in \cite{Diss}.
\par
In order to apply the {\sc Boltzmann}-like equations to behavioral changes
we have now to take the following specifications given by table 1
(cf. \cite{Diss}):
\par
\begin{table}[hbtp]
\begin{center}
\begin{tabular}{|l|l|}
\hline
STATISTICAL PHYSICS & BEHAVIORAL MODELS \\
\hline
\hline
system & population \\
\hline
subsystems & individuals \\
\hline
states & behaviors (concerning a \\
       & special topic of interest) \\
\hline
types of subsystems & types of behavior \\
                    & (subpopulations) \\
\hline
transitions & behavioral changes \\
\hline
interactions & imitative processes, \\
             & avoidance processes \\
\hline
fluctuations & ``freedom of will'' \\
\hline
\end{tabular}
\end{center}
\caption{Specification of the notions used in statistical physics for
an application to behavioral models.}
\end{table}
It is possible to generalize the resulting 
behavioral model to simultaneous
interactions of an arbitrary number of individuals (i.e., higher order
interactions) \cite{Helb,Diss}. 
However, in most cases behavioral changes are dominated
by pair interactions. Many of the phenomena occuring
in social interaction processes can already be understood by the discussion
of pair interactions.

\subsection{The form of the transition rates} \label{s2.1}

In the following we have to find a concrete form of the effective
transition rates\alpheqn{EFF}
\begin{equation}
 w^a(\vec{x}'|\vec{x};t) = w_a(\vec{x}'|\vec{x};t)
+ \sum_{b=1}^A \sum_{\weg{y}\in \Omega} \sum_{\weg{y}'\in \Omega} W_{ab}
(\vec{x}'|\vec{x},\vec{y};t) P_b(\vec{y},t) \, ,
\end{equation}
\begin{equation}
 W_{ab}(\vec{x}'|\vec{x},\vec{y};t) := \sum_{\weg{y}'\in \Omega}
 w_{ab}(\vec{x}',\vec{y}'|\vec{x},\vec{y};t)
\end{equation}\reseteqn
(cf. (\ref{effrate}))
that is suitable for the description of behavioral changes.
$W_{ab}(\vec{x}'|\vec{x},\vec{y};t)$ is the rate of pair interactions
\begin{equation}
 \vec{x}' \: \stackrel{\mbox{$\,\vec{y}$}}{\longleftarrow} \: \vec{x} \, ,
\end{equation}
where an individual of subpopulation $a$ changes the behavior from 
$\vec{x}$ to $\vec{x}'$ under the influence of an individual of 
subpopulation $b$ showing the behavior $\vec{y}$. There are only two
important kinds of social pair interactions:\alpheqn{ints}
\begin{eqnarray}
 \vec{x}' & \stackrel{\mbox{$\:\vec{x}'$}}{\longleftarrow} & \vec{x} 
 \qquad (\vec{x}' \ne \vec{x}) \\
 \vec{x}' & \stackrel{\mbox{$\,\vec{x}$}}{\longleftarrow} & \vec{x} 
 \qquad (\vec{x}' \ne \vec{x}) \, .
\end{eqnarray}\reseteqn
Obviously, the interpretation of the above kinds of pair interactions
is the following:
\begin{itemize}
\item The interactions (\ref{ints}a) describe {\em imitative processes}
(processes of persuasion), that means, the tendency to take over the
behavior $\vec{x}'$ of another individual.
\item The interactions (\ref{ints}b) describe {\em avoidance processes}, where
an individual changes the behavior when meeting another individual
showing the same behavior $\vec{x}$. 
Processes of this kind are known as aversive
behavior, defiant behavior or snob effect.  
\end{itemize}
The corresponding transition rates are of the general form\alpheqn{totrate}
\begin{eqnarray}
 W_{ab}(\vec{x}'|\vec{x},\vec{y};t) &:=&
 \nu_{ab}(t) R_{ab}^1(\vec{x}'|\vec{x};t) \delta_{\weg{y}\weg{x}'} \\
 &+& \nu_{ab}(t) R_{ab}^2(\vec{x}'|\vec{x};t) \delta_{\weg{y}\weg{x}} \, ,
\end{eqnarray}\reseteqn
where the term (\ref{totrate}a) describes imitative processes and the
term (\ref{totrate}b) describes avoidance processes. $\delta_{\weg{y}\weg{x}}$
has the meaning of the {\sc Kronecker} function. 
By inserting (\ref{totrate}) into (\ref{EFF})
we arrive at the following general form of the effective transition 
rates:\alpheqn{concrates}
\begin{equation}
 w^a(\vec{x}'|\vec{x};t) := \nu_a(t) R_a(\vec{x}'|\vec{x};t)
 + \sum_{b=1}^A \nu_{ab}(t) 
 \Big[ R_{ab}^1(\vec{x}'|\vec{x};t) P_b(\vec{x}',t) 
 + R_{ab}^2(\vec{x}'|\vec{x};t) P_b(\vec{x},t) \Big] \, .
\end{equation}
For behavioral models one often assumes
\begin{equation}
 R_{ab}^k(\vec{x}'|\vec{x};t) := f_{ab}^k(t) R^a(\vec{x}'|\vec{x};t) \, .
\end{equation}\reseteqn
In (\ref{concrates}),
\begin{itemize}
\item $\nu_a(t)$ is a measure for the rate of spontaneous 
(or externally induced) behavioral changes
within subpopulation $a$.
\item $R_a(\vec{x}'|\vec{x};t)$ [resp. $R^a(\vec{x}'|\vec{x};t)$] is the
{\em readiness} for an individual of subpopulation $a$ to change the behavior
from $\vec{x}$ to $\vec{x}'$ spontaneously [resp. in pair interactions].
\item $\nu_{ab}(t)\equiv N_b \, 
\widetilde{\nu}_{ab}(t)$ is the {\em interaction
rate} of an individual of subpopulation $a$ with individuals of subpopulation
$b$.
\item $f_{ab}^1(t)$ is a measure for the frequency of 
imitative processes.
\item $f_{ab}^2(t)$ is a measure for the frequency of avoidance 
processes.
\end{itemize}
A more detailled discussion of the different kinds of interaction processes
and of {\em ansatz} (\ref{concrates}) is given in 
\cite{Diss,Hel92}.
\par
For $R^a(\vec{x}'|\vec{x};t)$ we take the quite general form\alpheqn{Util}
\begin{equation}
 R^a(\vec{x}'|\vec{x};t) = \frac{\mbox{e}^{U^a(\weg{x}',t) - U^a(\weg{x},t)}}
 {D_a(\vec{x}',\vec{x};t)} 
 \label{util}
\end{equation}
with
\begin{displaymath}
 D_a(\vec{x}',\vec{x};t) = D_a(\vec{x},\vec{x}';t) > 0 
\end{displaymath}
(cf. \cite{Weid2,Diss}). Then, the readiness $R^a(\vec{x}'|\vec{x};t)$
for an individual of subpopulation $a$ to change the behavior from $\vec{x}$
to $\vec{x}'$ will be the greater, 
\begin{itemize}
\item the greater the {\em difference} of the
{\em utilities} $U^a(.,t)$ of behaviors $\vec{x}'$ and $\vec{x}$ is, 
\item the smaller the {\em incompatibility 
(``distance'')} $D_a(\vec{x}',\vec{x};t)$
between the behaviors $\vec{x}$ and $\vec{x}'$ is. 
\end{itemize}
Similar to (\ref{util}) we use 
\begin{equation}
 R_a(\vec{x}'|\vec{x};t) = \frac{\mbox{e}^{U_a(\weg{x}',t) - U_a(\weg{x},t)}}
 {D_a(\vec{x}',\vec{x};t)} \, ,
 \label{util2}
\end{equation}\reseteqn
and, therefore, allow the utility function $U_a(\vec{x},t)$ for spontaneous
(or externally induced)
behavioral changes to differ from the utility function $U^a(\vec{x},t)$
for behavioral changes in pair interactions.
{\em Ansatz} (\ref{Util}) is related to the {\em multinomial 
logit model} \cite{DoFa75,OrWi90}, 
and assumes {\em utility maximization} with 
incomplete information about the exact
utility of a behavioral change from $\vec{x}$ to $\vec{x}'$, which
is, therefore, estimated and stochastically varying (cf. \cite{Diss}). 
\par
Computer simulations of the {\sc Boltzmann}-like
equations (\ref{Boltz}), (\ref{concrates}), (\ref{Util}) 
are discussed and illustrated in 
\cite{Diss,He92,Hel92} (cf. also sect. \ref{s4}).

\subsection{Special fields of application in the social sciences} \label{s2.2}

The {\sc Boltzmann}-like equations (\ref{Boltz}), (\ref{concrates}) 
include a variety of special cases, which have become
very important in the social sciences: 
\begin{itemize}
\item The {\em logistic equation} \cite{Pea24,Ve45} describes limited growth
processes. Let us consider the situation of two behaviors $\vec{x}
\in \{ 1, 2 \}$ (i.e., $P_a(1,t) = 1 - P_a(2,t)$) 
and one subpopulation ($A=1$).
$\vec{x}=2$ may, for example, have the meaning to apply a certain
strategy, and $\vec{x} = 1$ not to do so. If only imitative
processes
\begin{equation}
 2 \stackrel{\mbox{2}}{\longleftarrow} 1
\end{equation}
and processes of spontaneous replacement
\begin{equation}
 1 \longleftarrow 2
\end{equation}
are considered, one arrives at the {\em logistic equation}
\begin{eqnarray}
 \frac{d}{dt} P_1(2,t) &=& - \nu_1(t) R_1(1|2;t)P_1(2,t)
 + \nu_{11}(t)f_{11}^1(t)R^1(2|1;t)\Big( 1 - P_1(2,t) \Big) P_1(2,t) \nonumber
 \\
 &\equiv & A(t) P_1(2,t) \Big( B(t) - P_1(2,t) \Big) \, .
\end{eqnarray}
\item The {\em gravity model} \cite{Zi46} describes processes of 
exchange between different places $\vec{x}$. It results for
$R_a(\vec{x}'|\vec{x};t) \equiv 0$, $f_{ab}^1(t) \equiv 1$,
$f_{ab}^2(t) \equiv 0$, and $A=1$:
\begin{equation}
 \frac{d}{dt} P(\vec{x},t) = \nu(t) \sum_{\weg{x}'\in \Omega} \left[ 
 \frac{\mbox{e}^{U(\weg{x},t) - U(\weg{x}',t)}}{D(\vec{x},\vec{x}')} 
 - \frac{\mbox{e}^{U(\weg{x}',t) - U(\weg{x},t)}}{D(\vec{x}',\vec{x})}
 \right] P(\vec{x},t)P(\vec{x}',t) \, .
\end{equation}
Here, we have dropped the index $a$ because of $a=1$. $P(\vec{x},t)$ is the
probability of being at place $\vec{x}$. The absolute rate
of exchange from $\vec{x}$ to $\vec{x}'$ is proportional to the
probabilities $P(\vec{x},t)$ and $P(\vec{x}',t)$ at the places $\vec{x}$
and $\vec{x}'$. $D(\vec{x},\vec{x}')$ is often chosen as a function 
of the metric distance $\| \vec{x} - \vec{x}' \|$ between
$\vec{x}$ and $\vec{x}'$: $D(\vec{x},\vec{x}') \equiv
D(\| \vec{x} - \vec{x}' \|)$.
\item The {\em behavioral model} of {\sc Weidlich} and {\sc Haag}
\cite{WeHa83,We91,Weid2}
assumes spontaneous transitions due to {\em indirect interactions},
which are, for example, 
induced by the media (TV, radio, or newspapers). We obtain
this model for $f_{ab}^1(t) \equiv 0 \equiv f_{ab}^2(t)$ and
\begin{equation}
 U_a(\vec{x},t) := \delta_a(\vec{x},t) 
 + \sum_{b=1}^A \kappa_{ab} \, P_b(\vec{x},t)
 \, . \label{kappa}
\end{equation}
$\delta_a(\vec{x},t)$ is the {\em preference} of subpopulation $a$ for 
behavior $\vec{x}$. $\kappa_{ab}$ are {\em coupling parameters} describing the
influence of the behaviorial distribution 
within subpopulation~$b$ on the behavior of
subpoplation $a$. For $\kappa_{ab} > 0$, $\kappa_{ab}$ reflects the
{\em social pressure} of behavioral majorities.
\item The {\em game dynamical equations} 
\cite{HoSi88,Mue90,Diss,He93} result for
$f_{ab}^1(t) \equiv \delta_{ab}$, $f_{ab}^2(t) \equiv 0$, and
\begin{equation}
 R^a(\vec{x}'|\vec{x};t) := \max \Big( E_a(\vec{x}',t) - E_a(\vec{x},t) , 0
 \Big) 
\end{equation}
(cf. \cite{Diss,Hel92}). 
Their explicit form 
is\alpheqn{game}
\begin{eqnarray}
 \frac{d}{dt} P_a(\vec{x},t) &=& \sum_{\weg{x}'\in \Omega} \Big[
 w_a(\vec{x}|\vec{x}';t)P_a(\vec{x}',t) - w_a(\vec{x}'|\vec{x};t)P_a(\vec{x},t)
 \Big] \\
 &+& \nu_{aa}(t) P_a(\vec{x},t) \Big[ E_a(\vec{x},t) - \langle E_a \rangle
 \Big] \, .
\end{eqnarray}\reseteqn
Whereas (\ref{game}a) again describes spontaneous behavioral changes
({\em ``mutations''}, innovations), 
(\ref{game}b) reflects competition processes
leading to a {\em ``selection''} of behaviors 
with a {\em success} $E_a(\vec{x},t)$
that exceeds the {\em average success}
\begin{equation}
 \langle E_a \rangle := \sum_{\weg{x}'\in \Omega} 
 E_a(\vec{x}',t) P_a(\vec{x}',t) \, .
\end{equation}
The success $E_a(\vec{x},t)$ is connected with the socalled {\em payoff
matrices} $\underline{A}_{ab} \equiv \Big( A_{ab}(\vec{x},\vec{y}) \Big)$ by
\begin{equation}
 E_a(\vec{x},t) := A_a(\vec{x}) 
 + \sum_{b=1}^A \sum_{\weg{y}\in \Omega} A_{ab}(\vec{x},\vec{y})
 P_b(\vec{y},t) 
\end{equation}
\cite{Diss,Hel92}. $A_a(\vec{x})$ means the success of behavior $\vec{x}$
with respect to the environment.
\par
Since the game dynamical equations (\ref{game}) 
agree with the {\em selection mutation equations} \cite{HoSi88}
they are not only a 
powerful tool in social sciences and economy 
\cite{Mue90,He93,Ax84,NeuMo44}, but also in
evolutionary biology \cite{Ei71,Fi30,EiSchu79,FeEb89}.
\end{itemize}

\section{The {\sc Boltzmann-Fokker-Planck} equations} \label{s3}

We shall now assume the set $\Omega$ of possible behaviors to build a
{\em continuous} space. The $n$ dimensions
of this space correspond to different characteristic
{\em aspects} of the considered behaviors.
In the continuous formulation, the sums in (\ref{Boltz}), (\ref{effrate})
have to be replaced by integrals:\alpheqn{Cont}
\begin{eqnarray}
 \frac{d}{dt} P_a(\vec{x},t) &=& \int\limits_\Omega d^n x' \, \Big[
 w^a(\vec{x}|\vec{x}';t) P_a(\vec{x}',t) 
 - w^a(\vec{x}'|\vec{x};t) P_a(\vec{x},t) \Big] \nonumber \\
&=& \int d^n x' \, \Big[
 w^a[\vec{x}'|\vec{x}-\vec{x}';t] P_a(\vec{x}-\vec{x}',t) 
 - w^a[\vec{x}'|\vec{x};t] P_a(\vec{x},t) \Big] \, ,
\end{eqnarray}
where
\begin{equation}
 w^a[\vec{x}'-\vec{x}|\vec{x};t] :=
 w^a(\vec{x}'|\vec{x};t) := w_a(\vec{x}'|\vec{x};t)
+ \sum_{b=1}^A \int\limits_\Omega d^n y 
\int\limits_\Omega d^n y' \, N_b \, \widetilde{w}_{ab}
(\vec{x}',\vec{y}'|\vec{x},\vec{y};t) P_b(\vec{y},t) \, .
\end{equation}\reseteqn
A reformulation of the {\sc Boltzmann}-like equations (\ref{Cont}) via  
a {\sc Kramers-Moyal} expansion \cite{Kr40,Mo49}
(second order {\sc Taylor} approximation) leads to 
a kind of {\em diffusion equations}:
the socalled {\sc Boltzmann-Fokker-Planck} {\em equations} 
\cite{Helb}\alpheqn{bfp}
\begin{equation}
 \frac{\partial}{\partial t}P_a(\vec{x},t) =
- \sum_{i=1}^n \frac{\partial}{\partial x_{i}} \Big[K_{a i}(\vec{x},t)
P_a(\vec{x},t)\Big]
+ \frac{1}{2} \sum_{i, j=1}^n 
\frac{\partial}{\partial x_{i}}\frac{\partial}
{\partial x_{j}} \Big[Q_{a i j}(\vec{x},t)
P_a(\vec{x},t)\Big] 
 \label{BFP}
\end{equation}
with the effective {\em drift coefficients}
\begin{equation}
K_{a i}(\vec{x},t) 
:= \int\limits_\Omega \!  d^n x' \, (x'_i - x_i) w^a(\vec{x}'|\vec{x};t) 
\label{effdr}
\end{equation}
and the effective {\em diffusion 
coefficients}\footnote{In \cite{Helb} 
the expression for 
$Q_{aij}(\vec{x},t)$ contains additional terms
due to another derivation of (\ref{bfp}).
However, they make no contributions, since they
result in vanishing surface integrals 
(cf. \cite{Diss}).}
\begin{equation}
Q_{a i j}(\vec{x},t) 
:= \displaystyle \int\limits_\Omega \!  d^n x' \, (x'_i - x_i) (x'_j - x_j)
 w^a(\vec{x}'|\vec{x};t) \, .
\label{effdiff}
\end{equation}\reseteqn
Whereas the drift coefficients $K_{ai}(\vec{x},t)$ govern the systematic
change of the distribution $P_a(\vec{x},t)$,
the diffusion coefficients $Q_{aij}(\vec{x},t)$ describe the
spread of the distribution $P_a(\vec{x},t)$ due to fluctuations
resulting from the individual variation of behavioral changes.
\par
For {\em ansatz} (\ref{concrates}), 
the effective drift and diffusion coefficients
can be splitted into contributions due to spontaneous 
(or externally induced) transitions
($k=0$), imitative processes ($k=1$), and avoidance processes 
($k=2$):\alpheqn{SUM}
\begin{equation}
 K_{ai}(\vec{x},t) = \sum_{k=0}^2 K_{ai}^k(\vec{x},t)\, , \qquad
 Q_{aij}(\vec{x},t) = \sum_{k=0}^2 Q_{aij}^k(\vec{x},t)\, ,
\label{split}
\end{equation}
where
\begin{eqnarray}
K_{ai}^0(\vec{x},t) &:=& \nu_a(t) \int \!  d^n x' \, (x'_i - x_i)
 R_a(\vec{x}'|\vec{x};t)\, , \nonumber \\
K_{ai}^1(\vec{x},t) &:=& \sum_{b=1}^A \nu_{ab}(t) f_{ab}^1(t) 
\int \!  d^n x' \, (x'_i - x_i)
 R^a(\vec{x}'|\vec{x};t)P_b(\vec{x}',t) \, , \nonumber \\
K_{ai}^2(\vec{x},t) &:=& \sum_{b=1}^A \nu_{ab}(t) f_{ab}^2(t) 
\int \!  d^n x' \, (x'_i - x_i)
 R^a(\vec{x}'|\vec{x};t)P_b(\vec{x},t) 
\label{splitforce}
\end{eqnarray}
and
\begin{eqnarray}
Q_{aij}^0(\vec{x},t) &:=& \nu_a(t) \int \!  d^n x' \, (x'_i-x_i) (x'_j-x_j)
 R_a(\vec{x}'|\vec{x};t)\, , \nonumber \\
Q_{aij}^1(\vec{x},t) &:=& \sum_{b=1}^A \nu_{ab}(t) f_{ab}^1(t) 
\int \!  d^n x' \, (x'_i - x_i)
(x'_j - x_j) R^a(\vec{x}'|\vec{x};t)P_b(\vec{x}',t) \, , \nonumber \\
Q_{aij}^2(\vec{x},t) &:=& \sum_{b=1}^A \nu_{ab}(t) f_{ab}^2(t) 
\int \!  d^n x' \, (x'_i - x_i)
(x'_j - x_j) R^a(\vec{x}'|\vec{x};t)P_b(\vec{x},t) \, .
\end{eqnarray}\reseteqn
The behavioral changes induced by the {\em environment} are included in
$K_{ai}^0(\vec{x},t)$ and $Q_{aij}^0(\vec{x},t)$.

\subsection{Social force and social field} \label{s3.1}

The {\sc Boltzmann-Fokker-Planck} equations (\ref{BFP}) are equivalent to 
the stochastic equations ({\sc Langevin} equations)\alpheqn{langevin}
\begin{equation}
 \frac{d x_i}{dt} = F_{ai}(\vec{x},t) 
 + \sum_{j=1}^n G_{aij}(\vec{x},t) \xi_j(t)
 \label{Langevin}
\end{equation}
with 
\begin{equation}
 K_{ai}(\vec{x},t) = F_{ai}(\vec{x},t) + \frac{1}{2} \sum_{j,k=1}^n
 \left[ \frac{\partial}{\partial x_k} G_{aij}(\vec{x},t) \right]
 G_{ajk}(\vec{x},t) 
 \label{Drift}
\end{equation}
and
\begin{equation}
 Q_{aij}(\vec{x},t) = \sum_{k=1}^n G_{aik}(\vec{x},t) G_{akj}(\vec{x},t)
 \label{Diff}
\end{equation}\reseteqn
(cf. \cite{Diss}). For an individual of subpopulation $a$
the vector $\vec{\zeta}_a(\vec{x},t)$ with the components
\begin{equation}
 \zeta_{ai}(\vec{x},t) = \sum_{j=1}^n G_{aij}(\vec{x},t) \xi_j(t)
\end{equation}
describes the contribution to the 
change of behavior $\vec{x}$ that is caused
by behavioral fluctuations $\vec{\xi}(t)$ (which are assumed
to be delta-correlated and {\sc Gauss}ian \cite{Diss}).   
Since the diffusion coefficients
$Q_{aij}(\vec{x},t)$ and the coefficients
$G_{aij}(\vec{x},t)$ are usually
small quantities, we have $F_{ai}(\vec{x},t) \approx
K_{ai}(\vec{x},t)$ (cf. (\ref{Drift})), 
and (\ref{Langevin}) can be put into the form
\begin{equation}
 \frac{d\vec{x}}{dt}  \approx \vec{K}_a(\vec{x},t) + \mbox{\em fluctuations.}
 \label{result}
\end{equation}
Whereas the fluctuation term describes individual behavioral variations, the
vectorial quantity $\vec{K}_a(\vec{x},t)$ 
drives the systematic change of the behavior $\vec{x}(t)$ of individuals of
subpopulation $a$. Therefore, it is justified to denote $\vec{K}_a(\vec{x},t)$
as {\em social force} acting on individuals of subpopulation $a$. 
\par
The social force influences the behavior of the individuals, but, conversely, 
due to interactions, the behavior of the individuals
also influences the social force via the behavioral distributions
$P_a(\vec{x},t)$ (cf. (\ref{Cont}b), (\ref{effdr})). 
That means, $K_a(\vec{x},t)$ is a
function of the social processes within the given population.
\par
Under the integrability conditions
\begin{equation}
 \frac{\partial}{\partial x_j} K_{ai}(\vec{x},t) =
 \frac{\partial}{\partial x_i} K_{aj}(\vec{x},t) \qquad
 \mbox{for all } i,j
 \label{intcond}
\end{equation}
there exists a time-dependent {\em potential}
\begin{equation}
 V_a(\vec{x},t) := - \int\limits^{\weg{x}} d\vec{x}' \cdot
 \vec{K}_a(\vec{x}',t) 
\, ,
 \label{potential}
\end{equation}
so that the social force is given by its gradient $\nabla$:
\begin{equation}
 \vec{K}_a(\vec{x},t) = - \nabla V_a(\vec{x},t)
\, .
 \label{gradient}
\end{equation}
The potential $V_a(\vec{x},t)$ can be understood as {\em social field}.
It reflects the social influences and interactions relevant for 
behavioral changes: the public opinion, trends, social norms, etc.

\subsection{Discussion of the concept of force} \label{s3.2}

Clearly, the social force is no force obeying the {\sc Newton}ian laws of
mechanics. Instead, the social force $\vec{K}_a(\vec{x},t)$ is a vectorial
quantity with the following properties:
\begin{itemize}
\item $\vec{K}_a(\vec{x},t)$ drives the temporal change $d\vec{x}/dt$ of
another vectorial quantity: the behavior $\vec{x}(t)$ of an individual
of subpopulation $a$.
\item The component
\begin{equation}
 \vec{K}_{ab}(\vec{x},t) := \nu_{ab}(t) \int\limits_\Omega d^n x' \, (\vec{x}'
 - \vec{x}) \Big[ f_{ab}^1(t) P_b(\vec{x}',t)
 + f_{ab}^2(t) P_b(\vec{x},t) \Big] R^a(\vec{x}'|\vec{x};t) 
\end{equation}
of the social force $\vec{K}_a(\vec{x},t)$
describes the reaction of subpopulation $a$ on the behavioral distribution
within subpopulation $b$ and usually differs from 
$\vec{K}_{ba}(\vec{x},t)$, which describes the influence of subpopulation $a$
on subpopulation $b$.
\item Neglecting fluctuations, the behavior $\vec{x}(t)$ does not change
if $\vec{K}_a(\vec{x},t)$ vanishes. 
$\vec{K}_a(\vec{x},t) = \mbox{\bf 0}$ corresponds
to an {\em extremum} of the social field $V_a(\vec{x},t)$, because it means
\begin{equation}
 \nabla V_a(\vec{x},t) = \mbox{\bf 0} 
\, .
\end{equation}
\end{itemize}
We can now formulate our results in the following form 
related to {\sc Lewin}'s {\em ``field theory''} \cite{Le51}:
\begin{itemize}
\item Let us assume that an individuals objective is to behave in an
optimal way with respect to the social field, that
means, he or she tends to a behavior corresponding to a {\em minimum} of the
social field.
\item If the behavior $\vec{x}$ does not agree with a minimum of the social
field $V_a(\vec{x},t)$ this evokes a {\em psychical tension (force)}
\begin{equation}
 \vec{K}_a(\vec{x},t) = - \nabla V_a(\vec{x},t) 
\end{equation}
that is given by the gradient of the social field $V_a(\vec{x},t)$.
\item The psychical tension $\vec{K}_a(\vec{x},t)$ is a vectorial quantity
that induces a behavioral change according to
\begin{equation}
 \frac{d\vec{x}}{dt} \approx \vec{K}_a(\vec{x},t) \, .
\label{accor}
\end{equation}
\item The behavioral change $d\vec{x}/dt$ drives the behavior $\vec{x}(t)$
towards a minimum $\vec{x}_a^*$ of the social field $V_a(\vec{x},t)$.
\item When the behavior has reached a minimum 
$\vec{x}_a^*$ of the social field $V_a(\vec{x},t)$, it holds
\begin{equation}
 \nabla V_a(\vec{x},t) = \mbox{\bf 0}
\end{equation}
and, therefore, $\vec{K}_a(\vec{x},t) = \mbox{\bf 0}$, 
that means, the psychical tension vanishes.
\item If the psychical tension $\vec{K}_a(\vec{x},t)$ vanishes, 
except for fluctuations
no behavioral changes take place---in accordance with (\ref{accor}). 
\end{itemize}
In the special case, where an individual's objective is the behavior
$\vec{x}_a^*$, one would expect behavioral changes according to
\begin{equation}
 \frac{d\vec{x}}{dt} \approx \gamma ( \vec{x}_a^* - \vec{x} ) \, ,
\end{equation}
which corresponds to a social field
\begin{equation}
 V_a(\vec{x},t) \approx \frac{\gamma}{2} (\vec{x}_a^* - \vec{x})^2
\end{equation}
with a minimum at $\vec{x}_a^*$. Examples for this case are discussed in
\cite{He91}.
\par
Note, that the social fields $V_a(\vec{x},t)$ of 
different subpopulations $a$ usually have 
{\em different} minima $\vec{x}_a^*$.
That means, individuals of different subpopulations $a$
will normally feel different psychical tensions
$\vec{K}_a(\vec{x},t)$. This shows the psychical
tension $\vec{K}_a(\vec{x},t)$ to be a {\em ``subjective''} quantity.

\section{Computer simulations} \label{s4}

\begin{figure}[htbp]
\parbox[b]{7.4cm}{
\epsfxsize=7.3cm 
\centerline{\rotate[r]{\hbox{\epsffile[28 28 570
556]{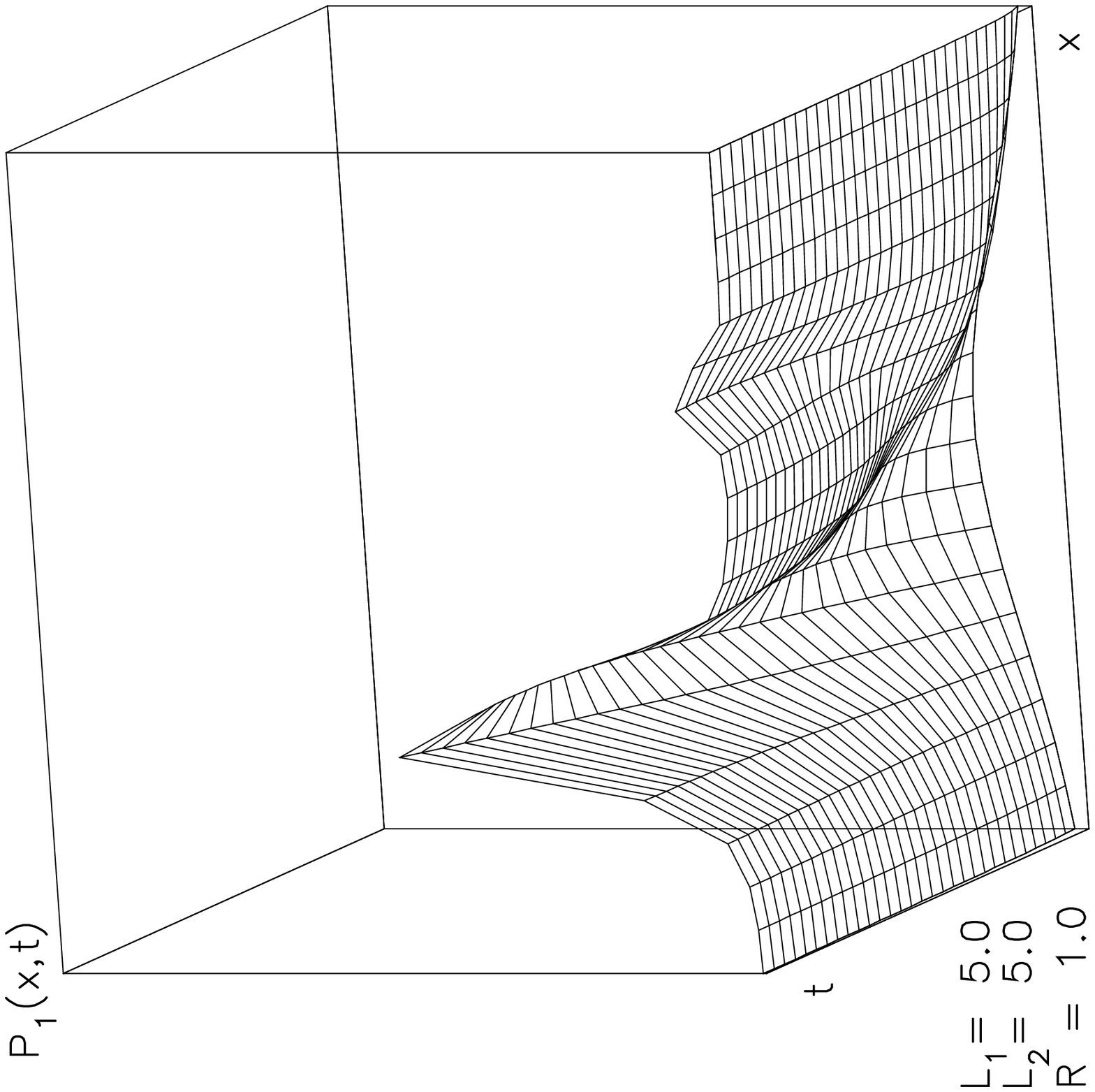}}}}
}\hfill
\parbox[b]{7.4cm}{
\epsfxsize=7.3cm 
\centerline{\rotate[r]{\hbox{\epsffile[28 28 570
556]{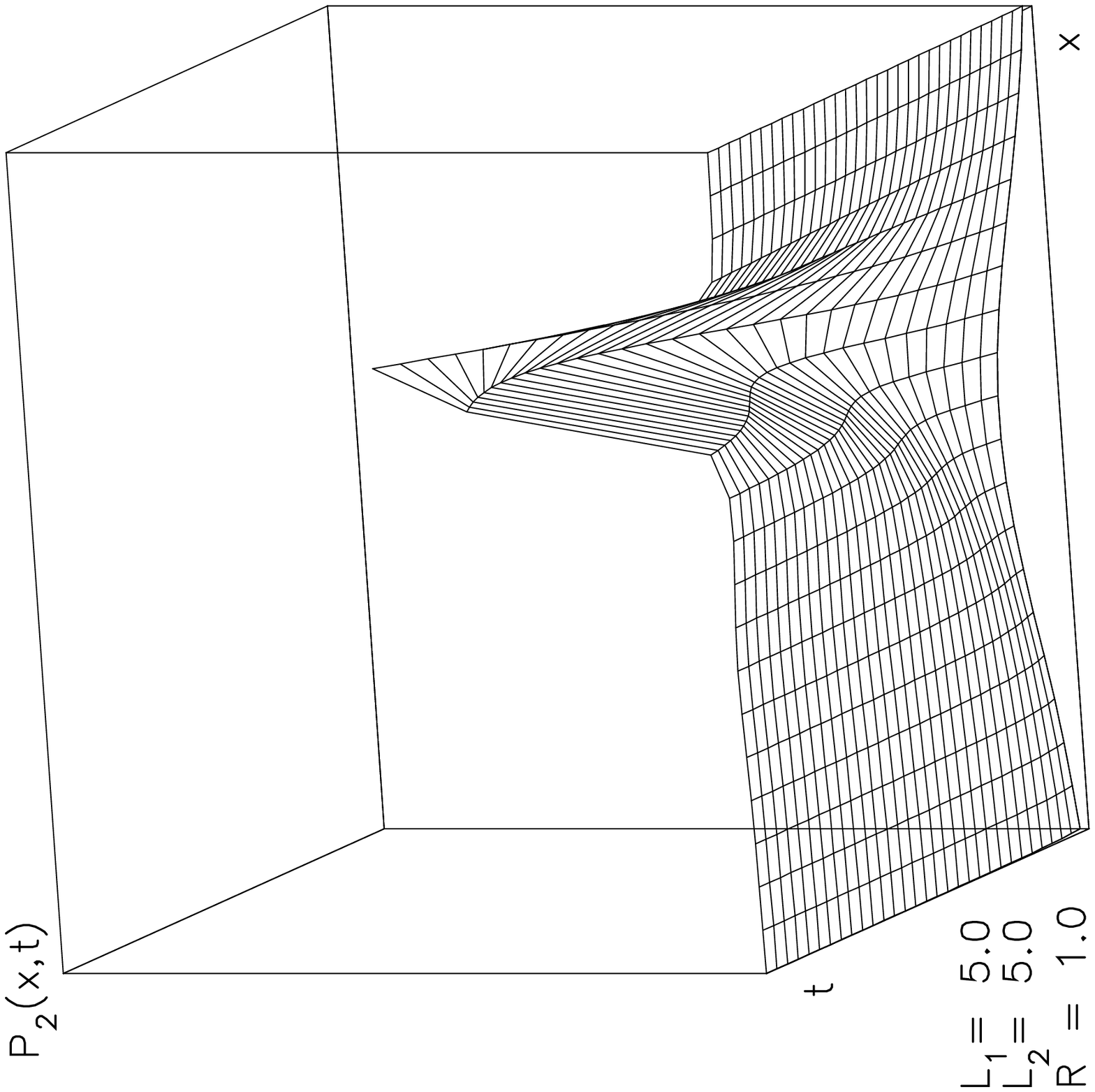}}}}
}  
\parbox[b]{7.4cm}{
\epsfxsize=7.3cm 
\centerline{\rotate[r]{\hbox{\epsffile[28 28 570
556]{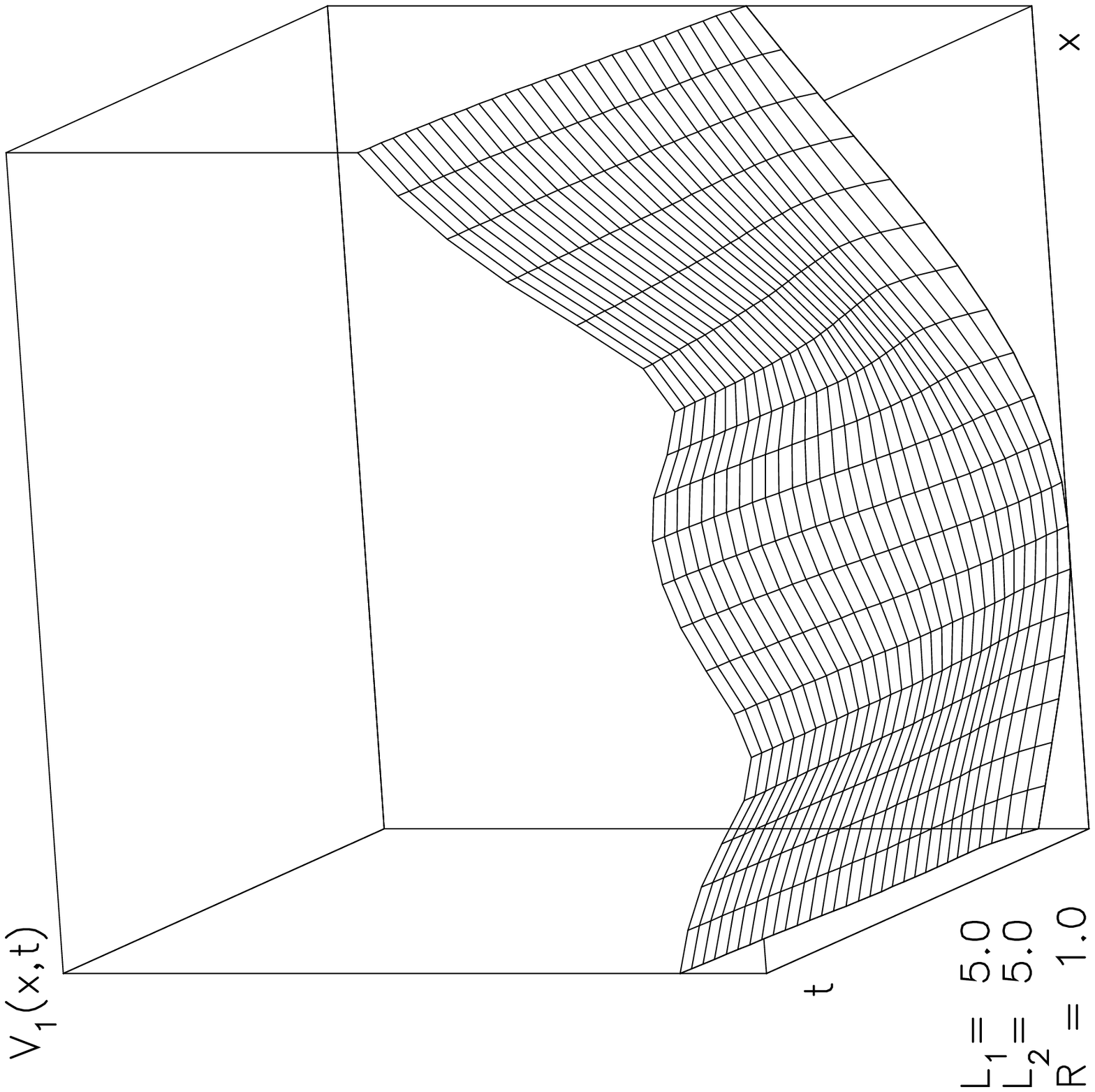}}}}
}\hfill
\parbox[b]{7.4cm}{
\epsfxsize=7.3cm 
\centerline{\rotate[r]{\hbox{\epsffile[28 28 570
556]{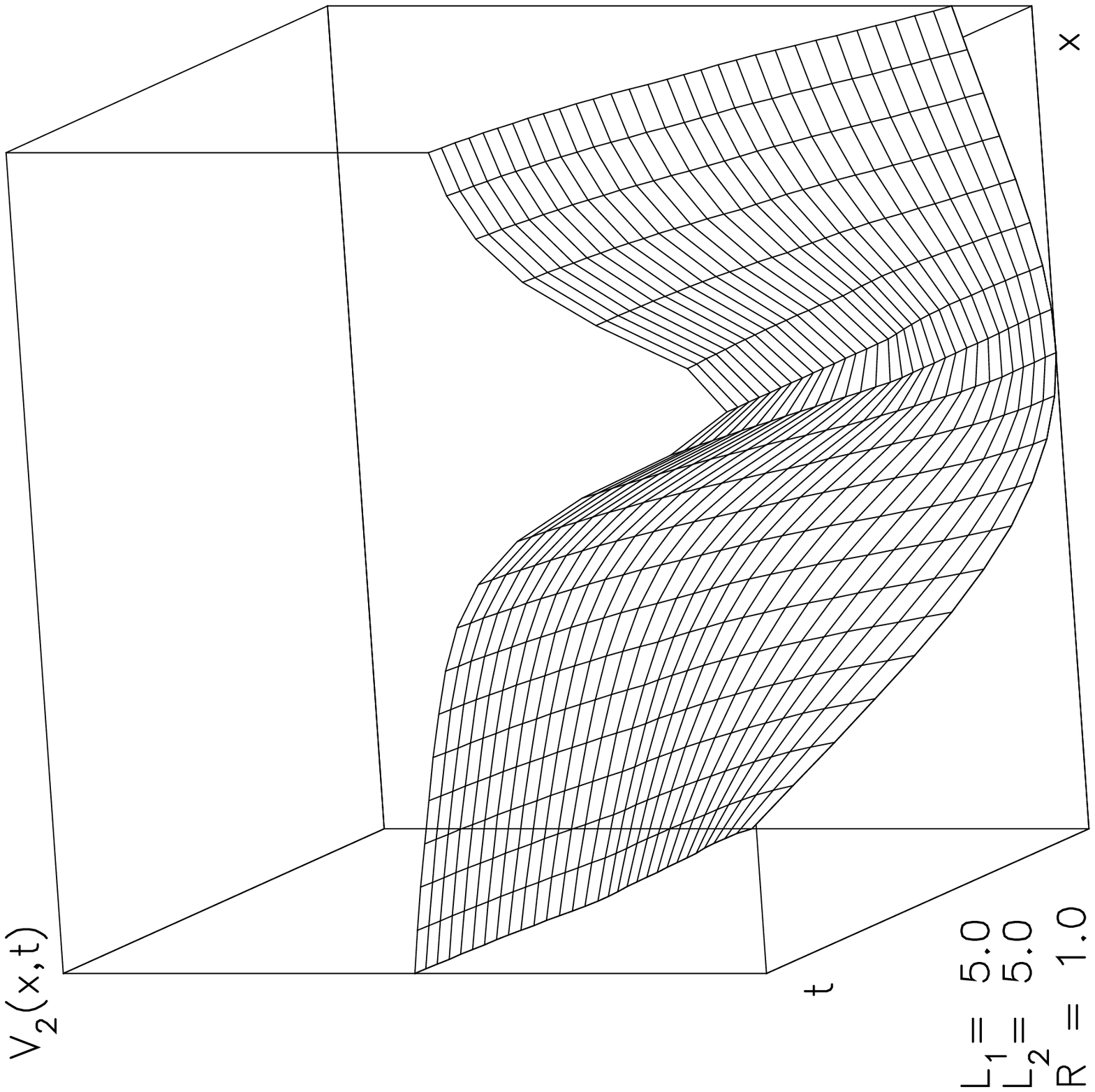}}}}
}
\parbox[b]{7.4cm}{
\epsfxsize=7.3cm 
\centerline{\rotate[r]{\hbox{\epsffile[28 28 570
556]{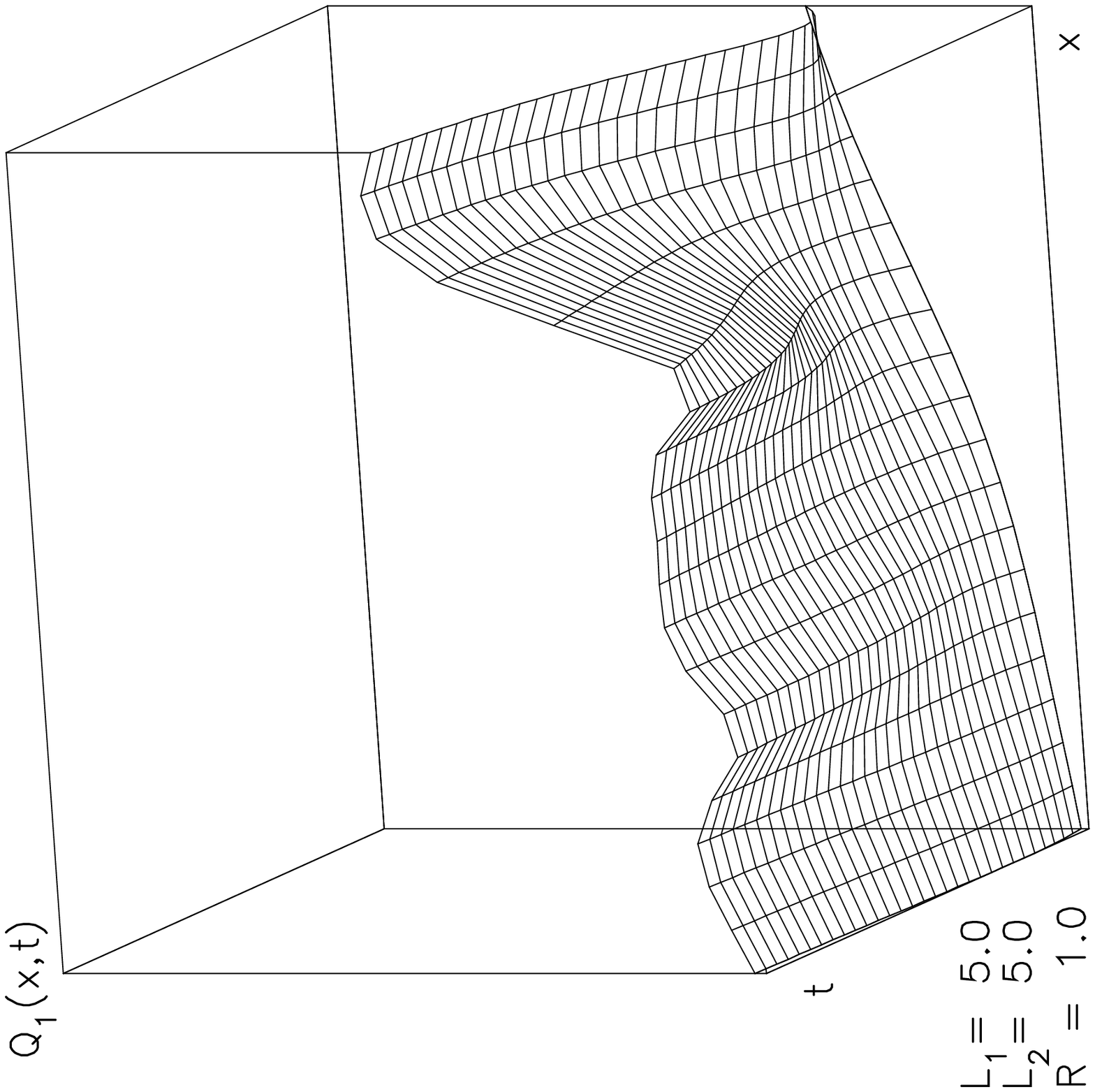}}}}
}\hfill
\parbox[b]{7.4cm}{
\epsfxsize=7.3cm 
\centerline{\rotate[r]{\hbox{\epsffile[28 28 570
556]{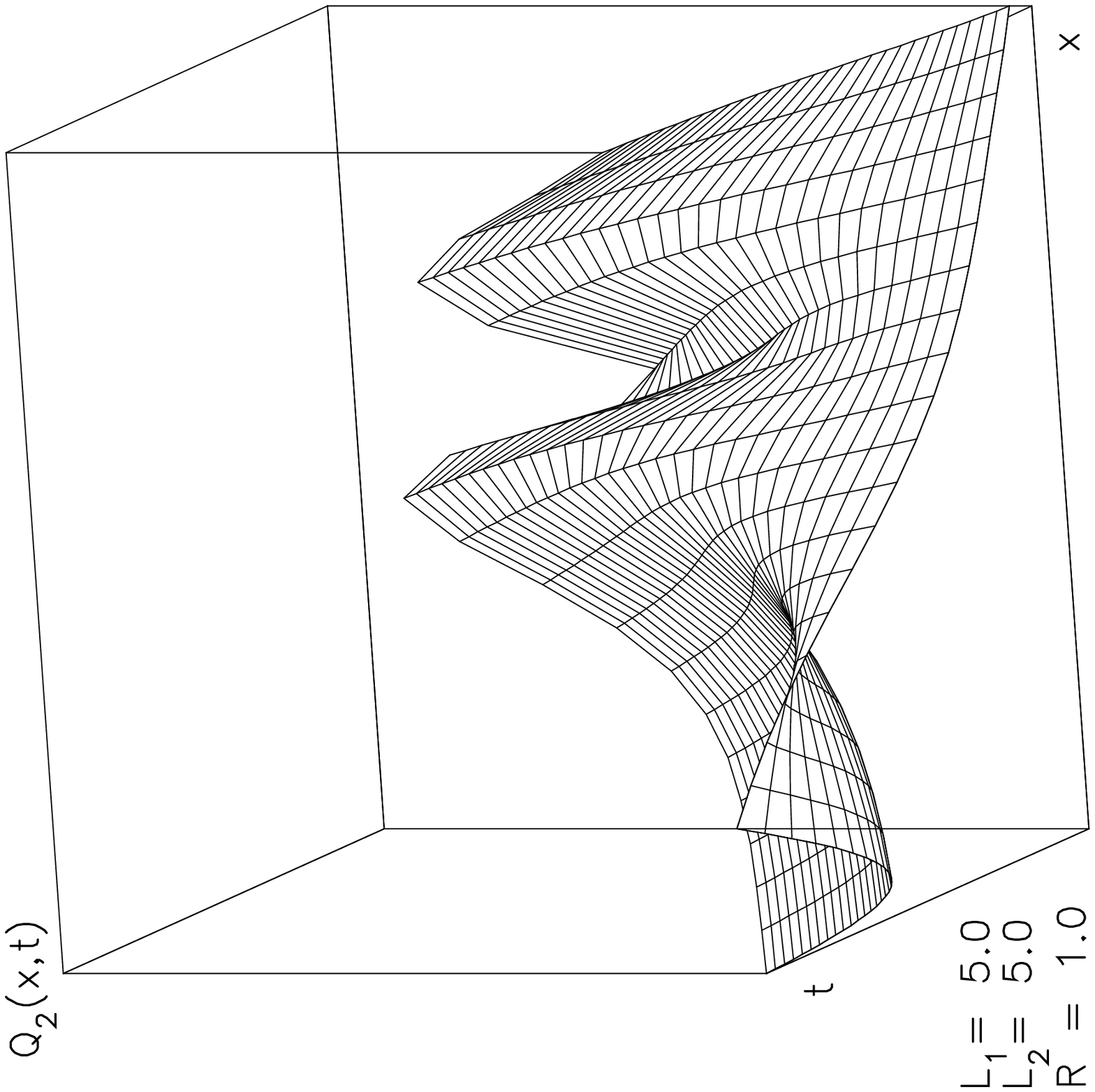}}}}
}
\parbox{15cm}{
\caption{Imitative processes in the case of one-sided sympathy and
low indifference $L_a$ with respect to behavioral changes.\label{fi1}}
}
\end{figure}

\begin{figure}[htbp]
\parbox[b]{7.4cm}{
\epsfxsize=7.3cm 
\centerline{\rotate[r]{\hbox{\epsffile[28 28 570
556]{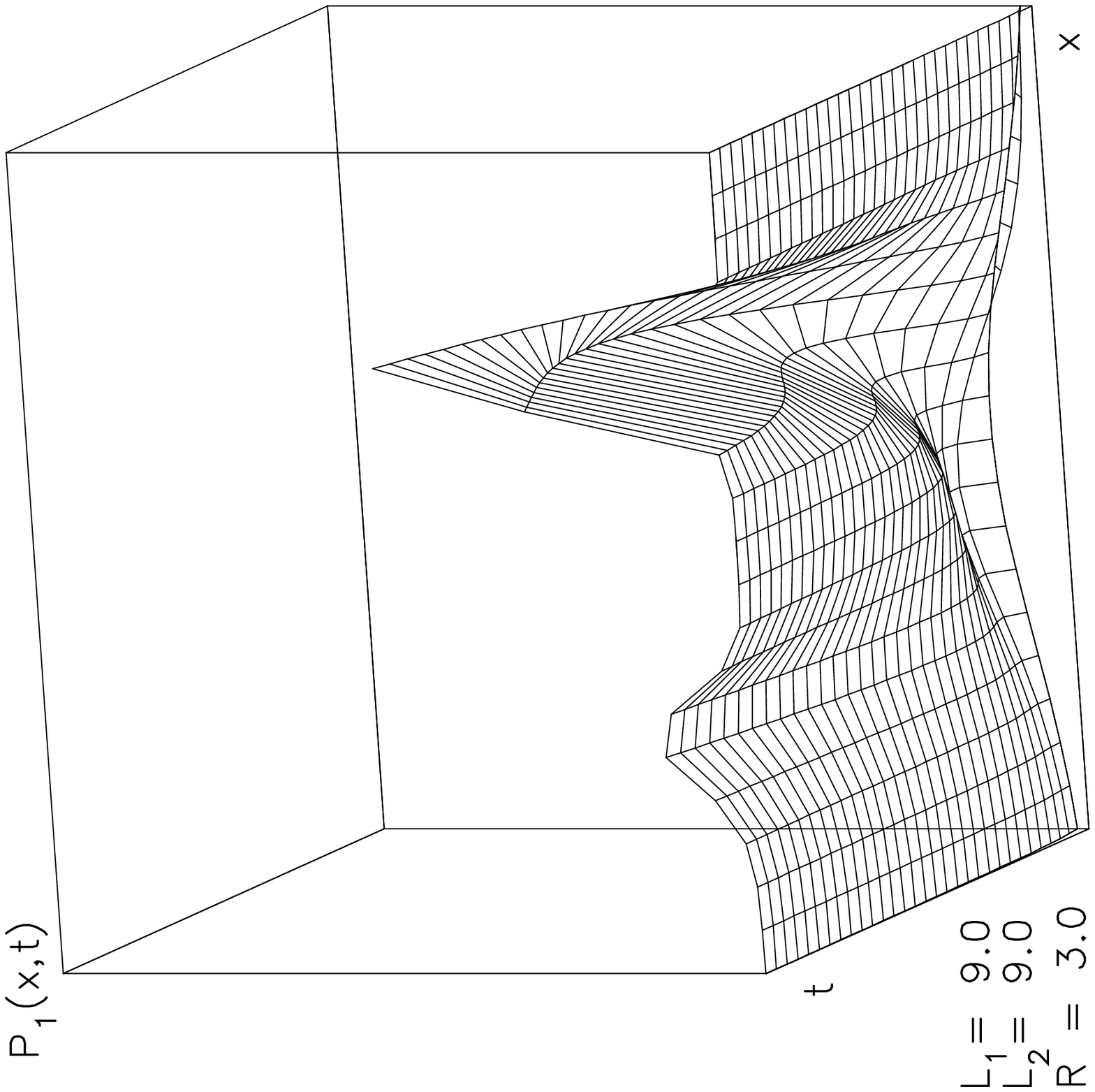}}}}
}\hfill
\parbox[b]{7.4cm}{
\epsfxsize=7.3cm 
\centerline{\rotate[r]{\hbox{\epsffile[28 28 570
556]{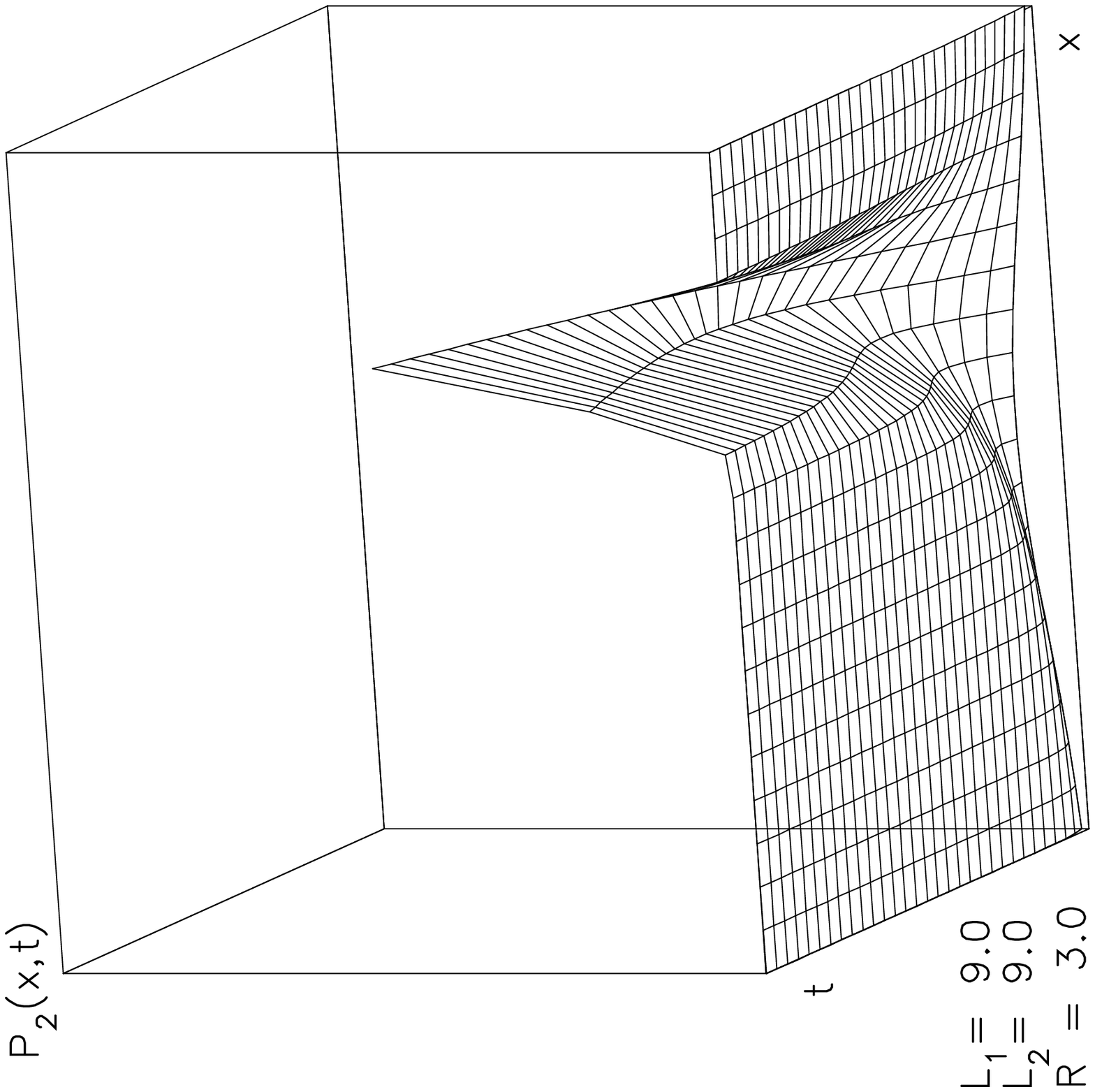}}}}
}
\parbox[b]{7.4cm}{
\epsfxsize=7.3cm 
\centerline{\rotate[r]{\hbox{\epsffile[28 28 570
556]{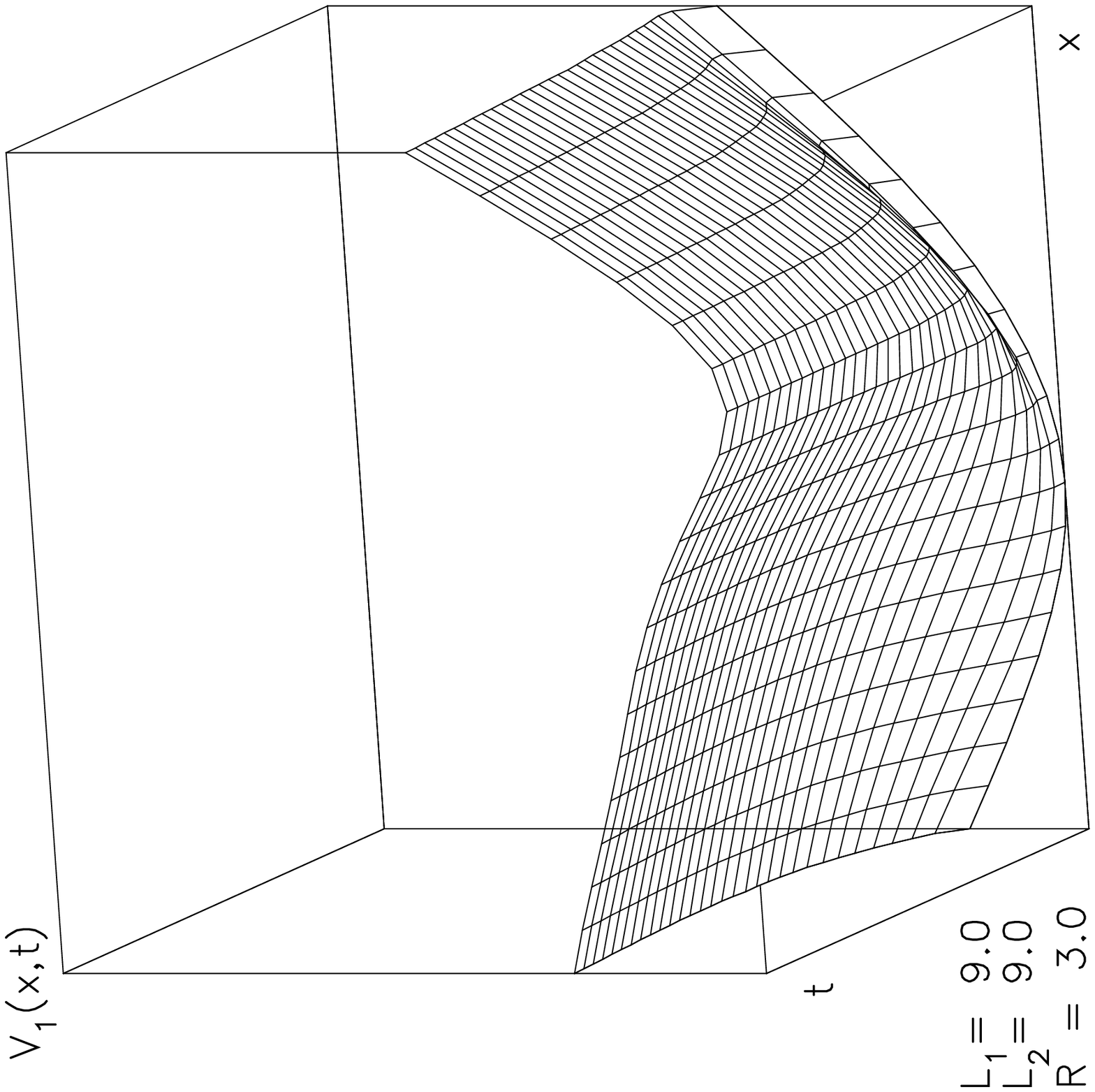}}}}
}\hfill
\parbox[b]{7.4cm}{
\epsfxsize=7.3cm 
\centerline{\rotate[r]{\hbox{\epsffile[28 28 570
556]{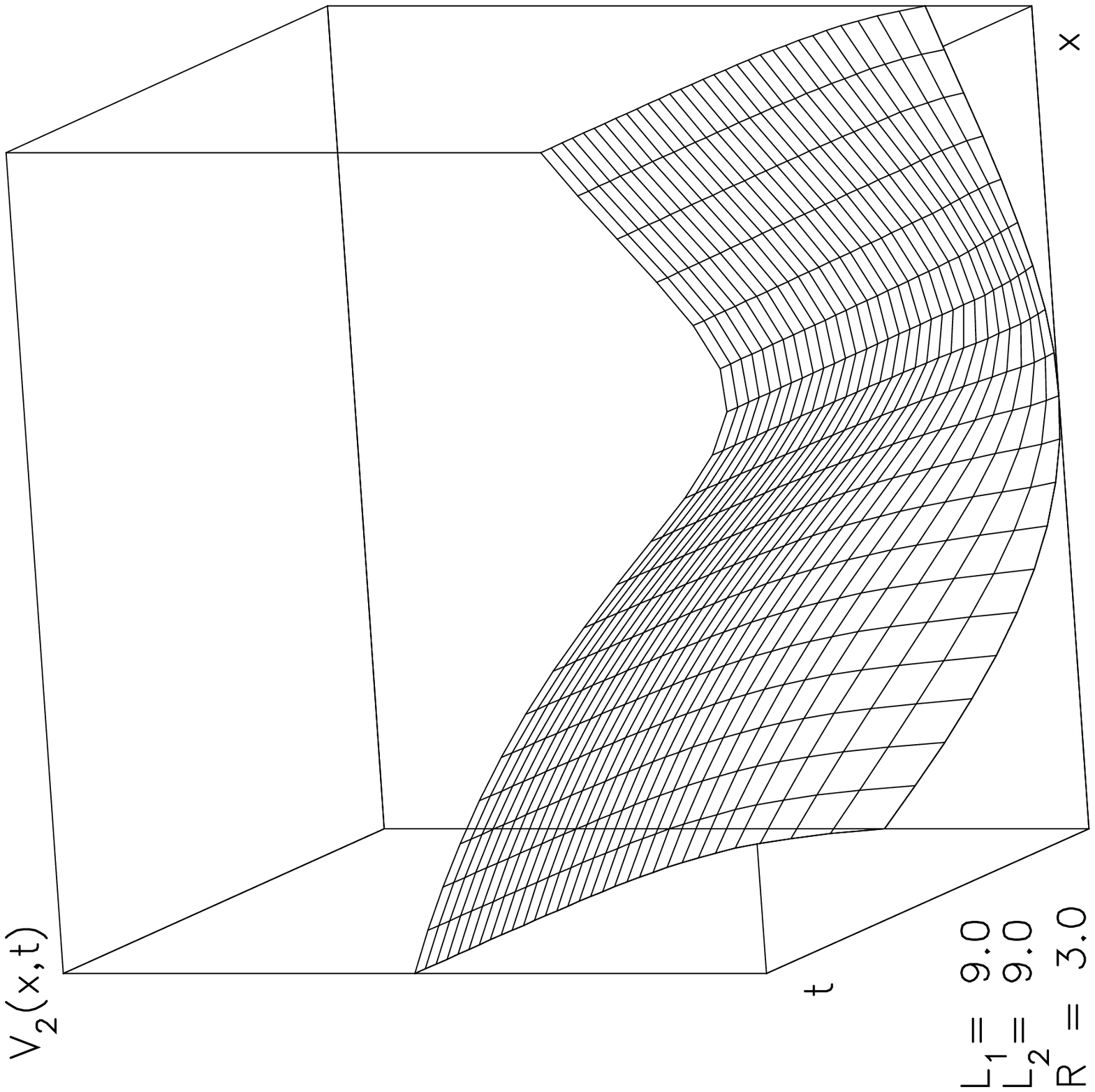}}}}
}
\parbox[b]{7.4cm}{
\epsfxsize=7.3cm 
\centerline{\rotate[r]{\hbox{\epsffile[28 28 570
556]{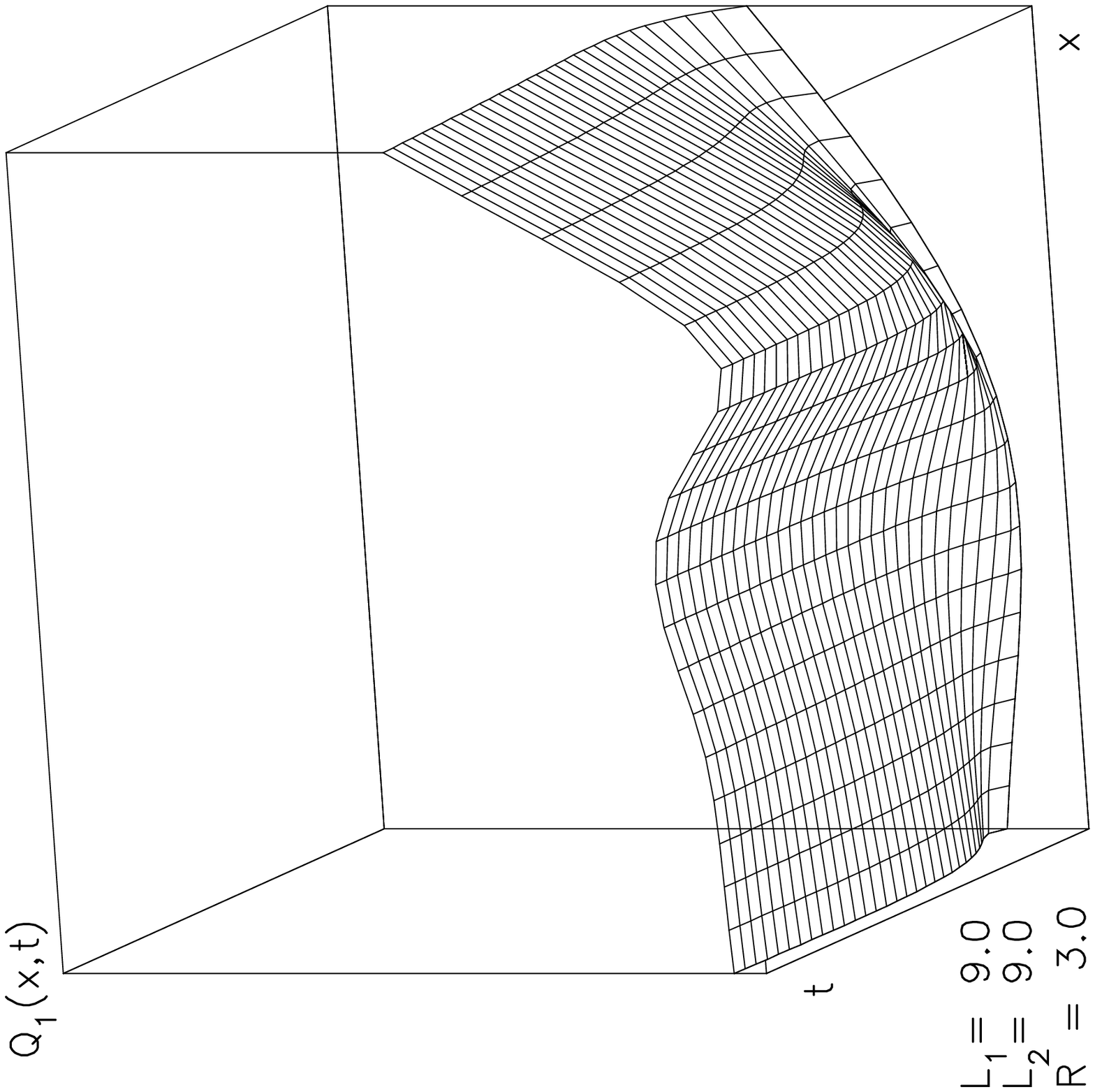}}}}
}\hfill
\parbox[b]{7.4cm}{
\epsfxsize=7.3cm 
\centerline{\rotate[r]{\hbox{\epsffile[28 28 570
556]{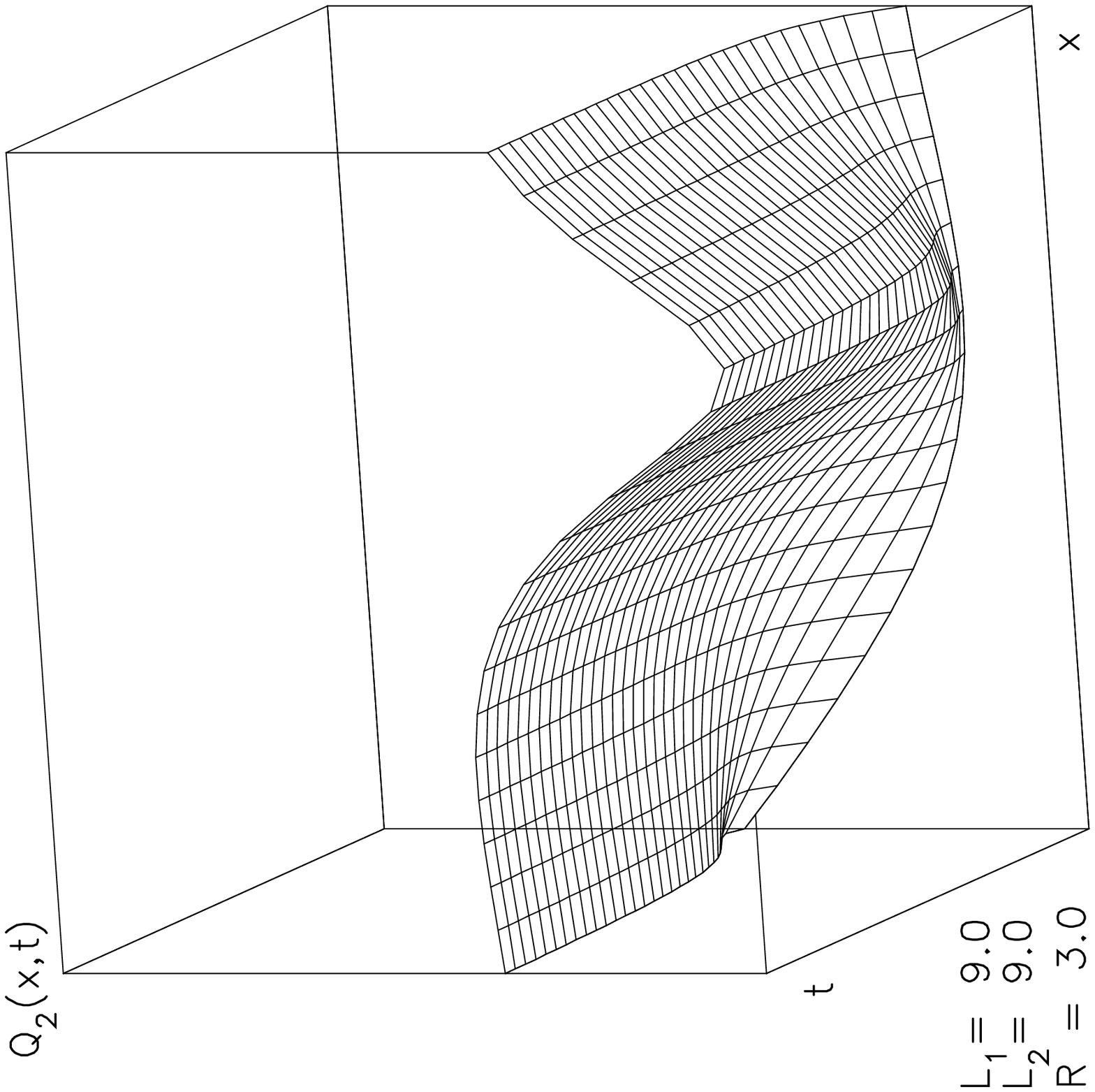}}}}
}
\parbox{15cm}{
\caption{As figure 1, but for high indifference $L_a$ with respect
to behavioral changes.\label{fi2}}
}
\end{figure}

\begin{figure}[htbp]
\parbox[b]{7.4cm}{
\epsfxsize=7.3cm 
\centerline{\rotate[r]{\hbox{\epsffile[28 28 570
556]{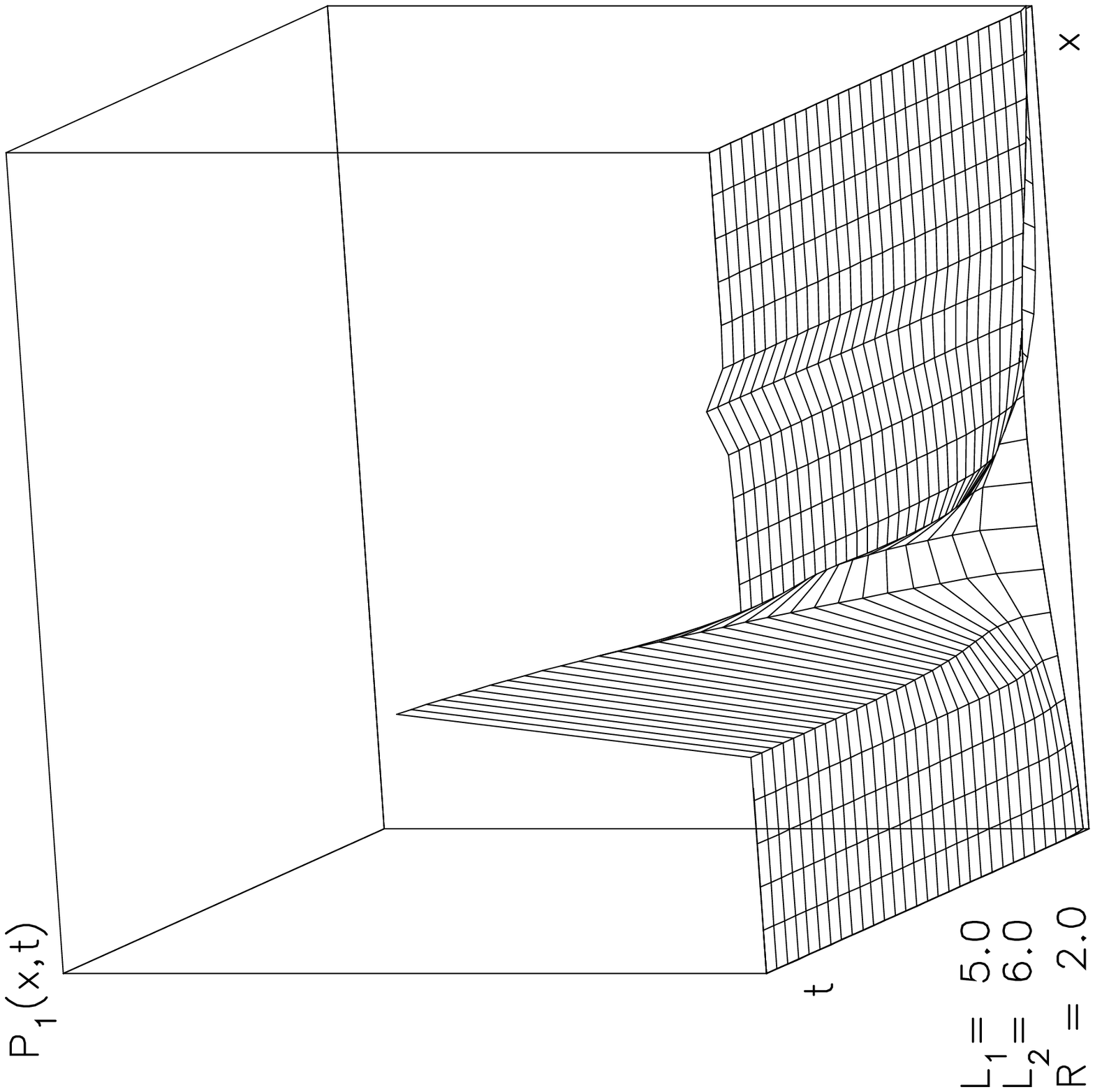}}}}
}\hfill
\parbox[b]{7.4cm}{
\epsfxsize=7.3cm 
\centerline{\rotate[r]{\hbox{\epsffile[28 28 570
556]{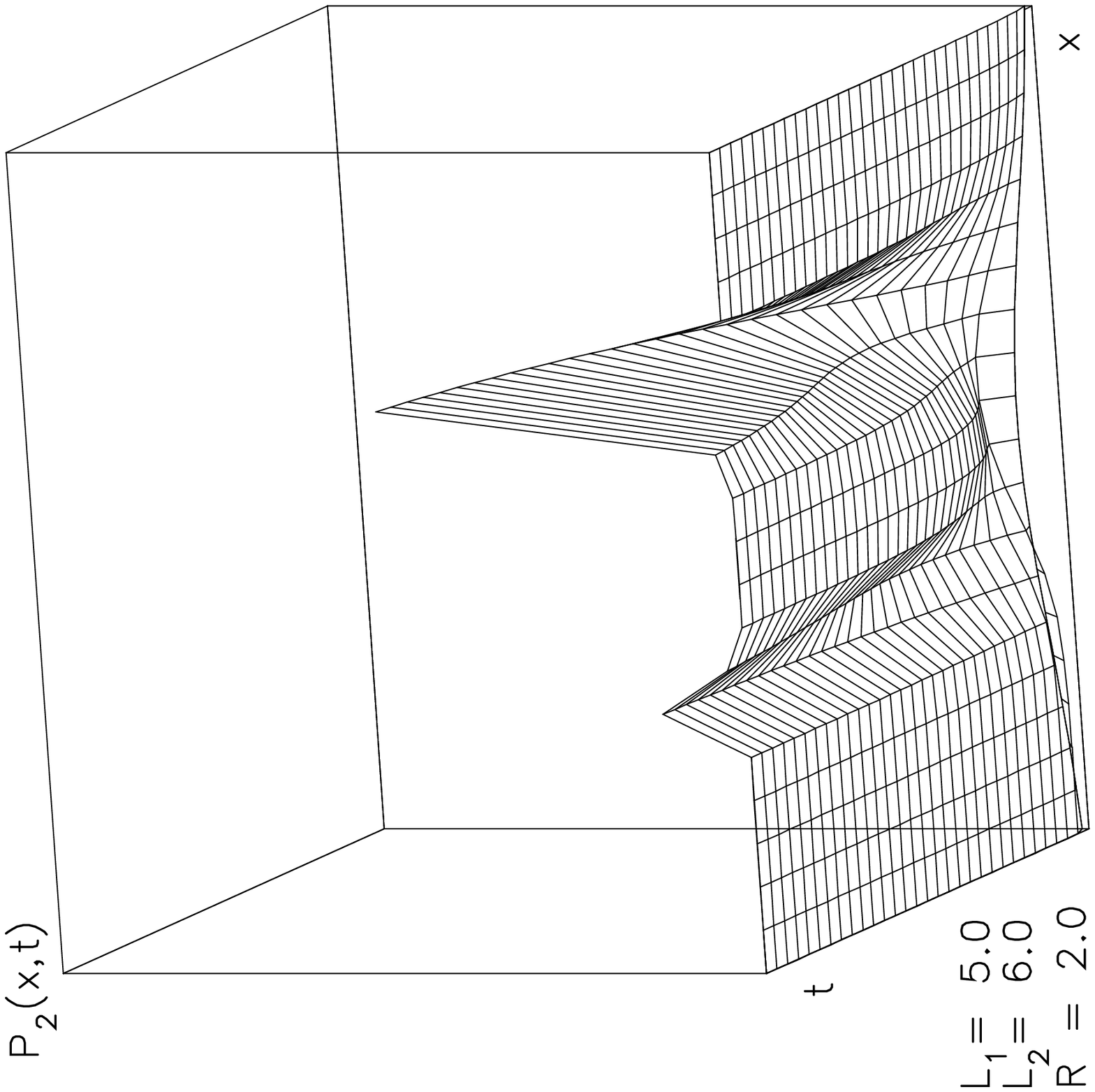}}}}
}
\parbox[b]{7.4cm}{
\epsfxsize=7.3cm 
\centerline{\rotate[r]{\hbox{\epsffile[28 28 570
556]{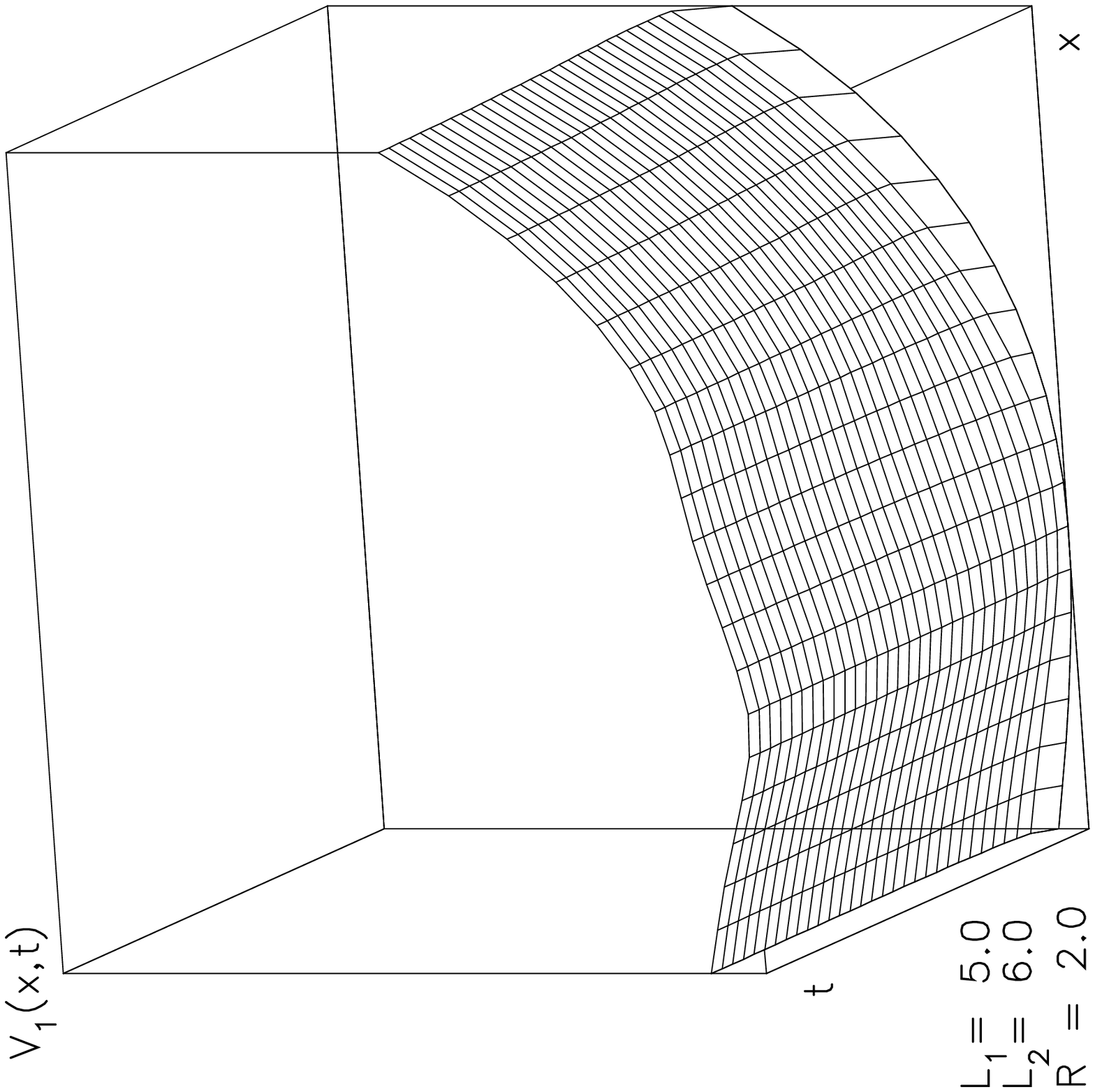}}}}
}\hfill
\parbox[b]{7.4cm}{
\epsfxsize=7.3cm 
\centerline{\rotate[r]{\hbox{\epsffile[28 28 570
556]{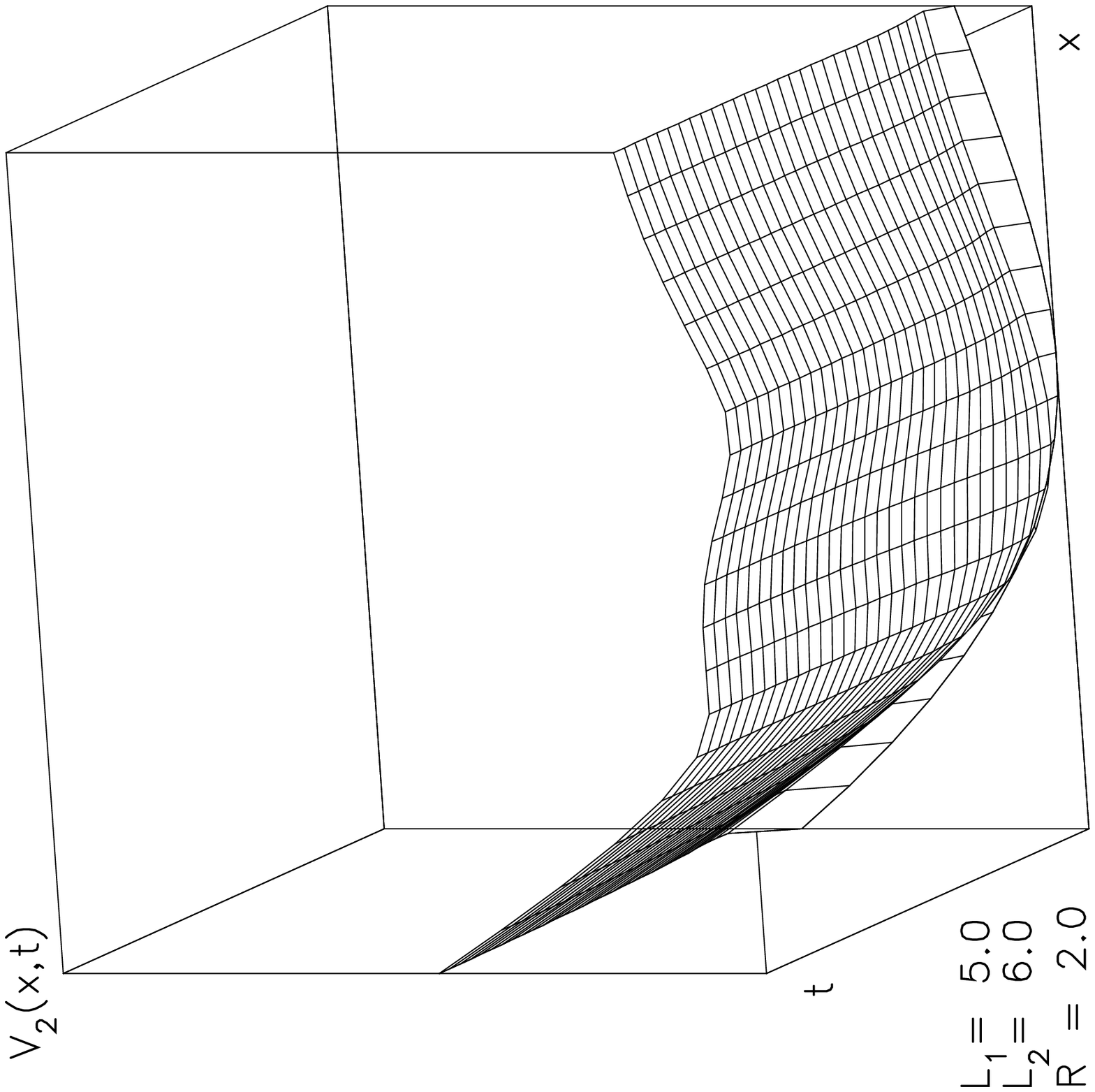}}}}
}
\parbox[b]{7.4cm}{
\epsfxsize=7.3cm 
\centerline{\rotate[r]{\hbox{\epsffile[28 28 570
556]{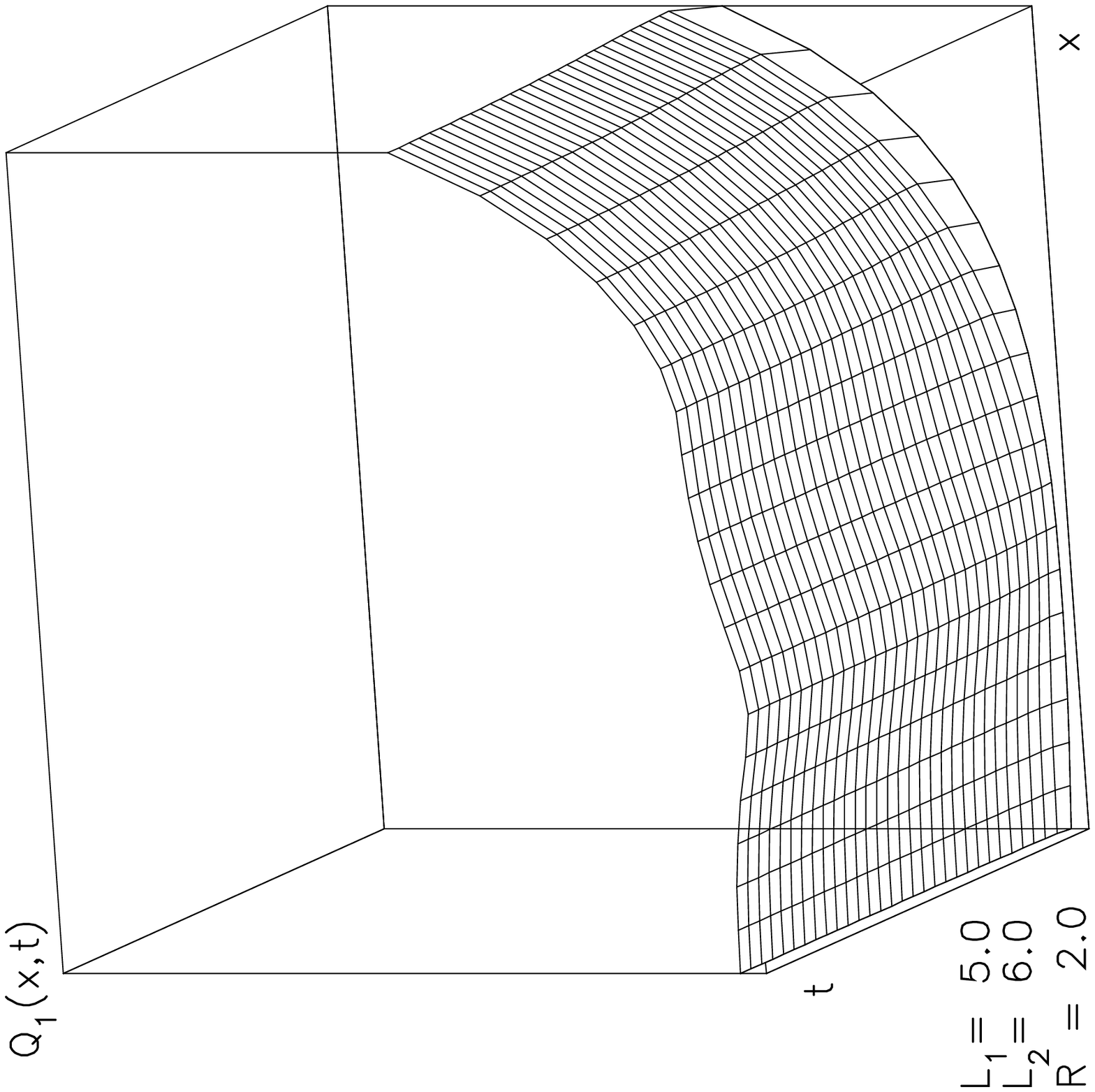}}}}
}\hfill
\parbox[b]{7.4cm}{
\epsfxsize=7.3cm 
\centerline{\rotate[r]{\hbox{\epsffile[28 28 570
556]{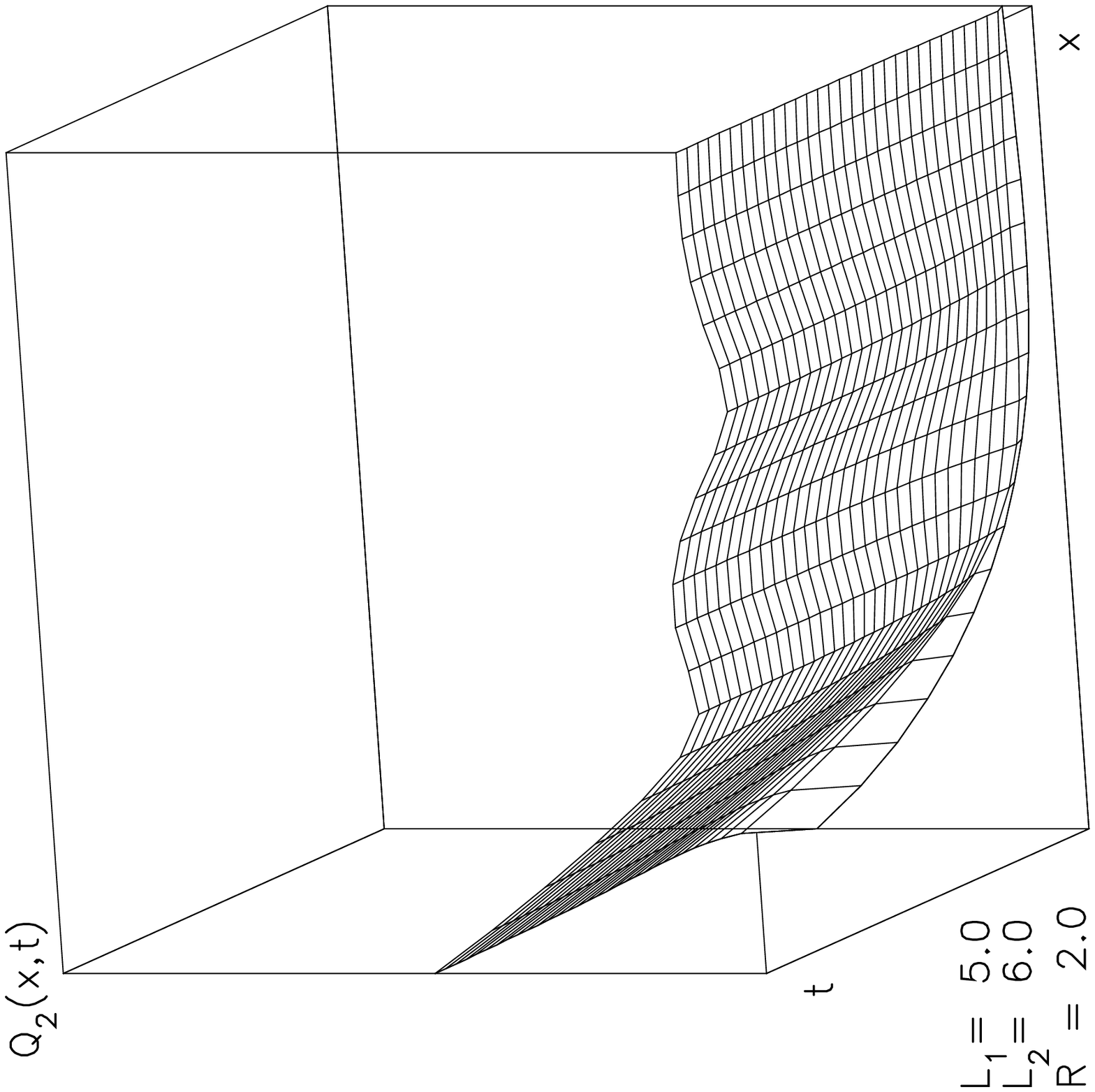}}}}
}
\parbox{15cm}{
\caption{Imitative processes for mutual sympathy and low indifference $L_a$
in both subpopulations.\label{fi3}}
}
\end{figure}

\begin{figure}[htbp]
\parbox[b]{7.4cm}{
\epsfxsize=7.3cm 
\centerline{\rotate[r]{\hbox{\epsffile[28 28 570
556]{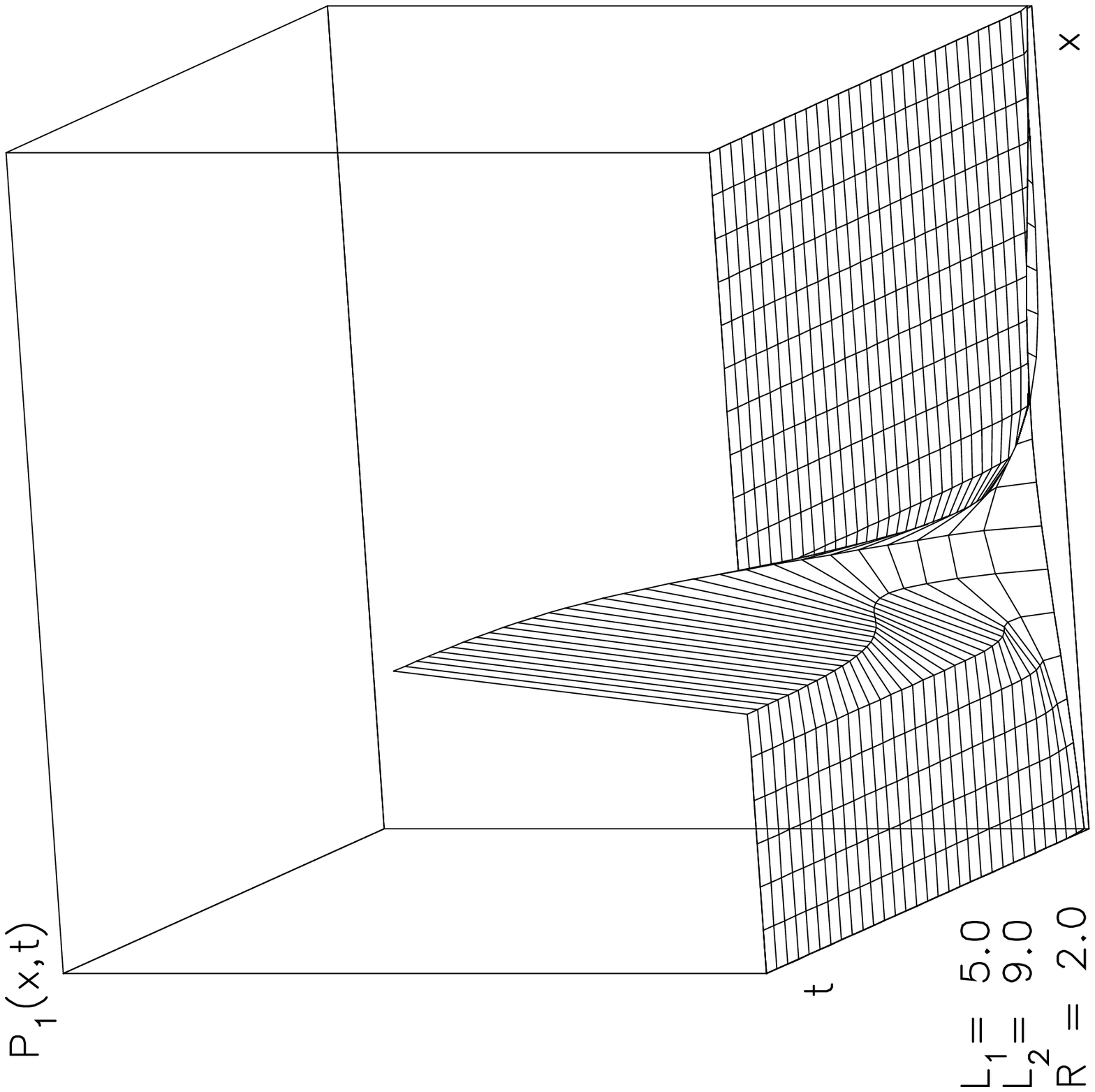}}}}
}\hfill
\parbox[b]{7.4cm}{
\epsfxsize=7.3cm 
\centerline{\rotate[r]{\hbox{\epsffile[28 28 570
556]{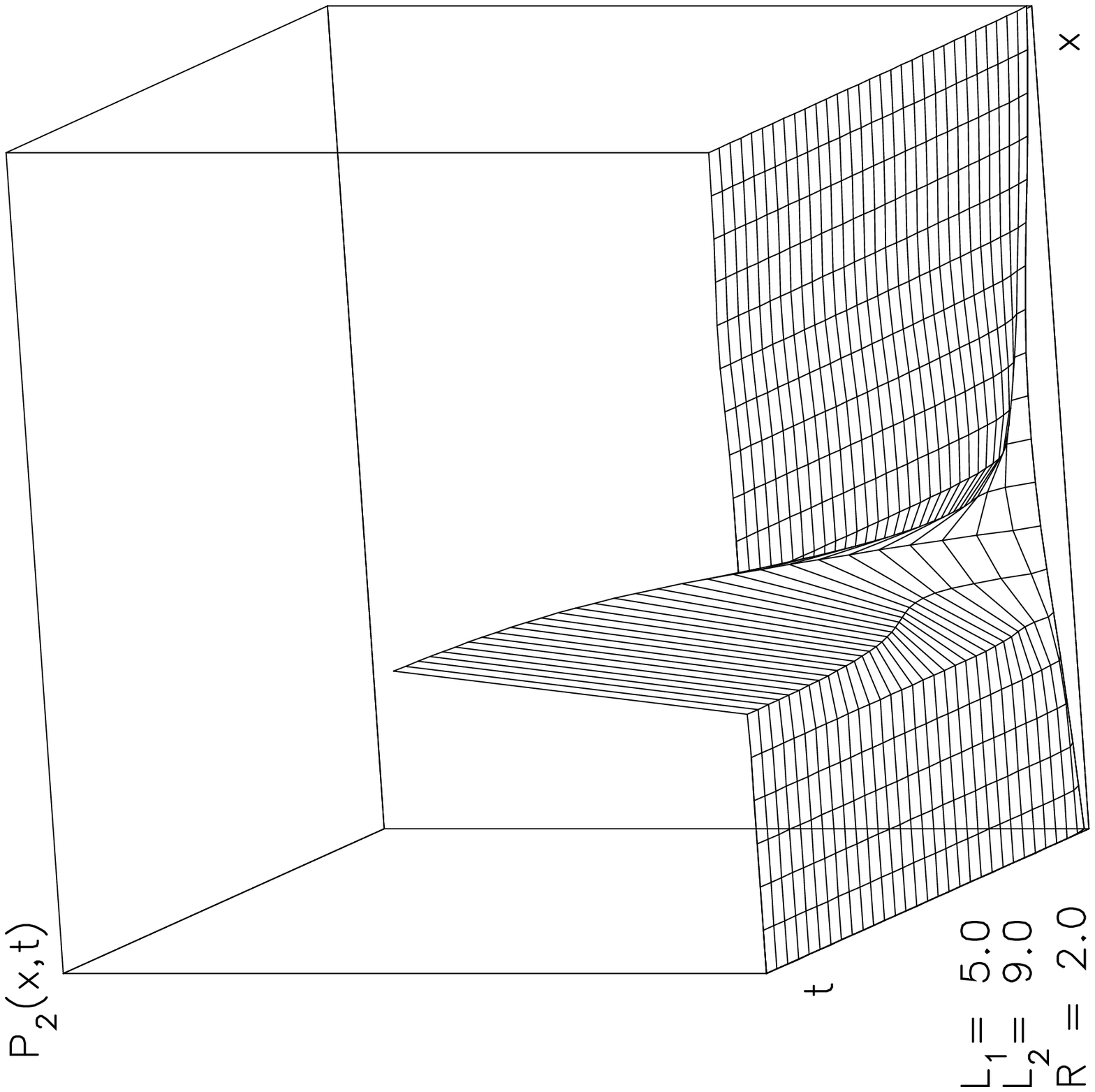}}}}
}
\parbox[b]{7.4cm}{
\epsfxsize=7.3cm 
\centerline{\rotate[r]{\hbox{\epsffile[28 28 570
556]{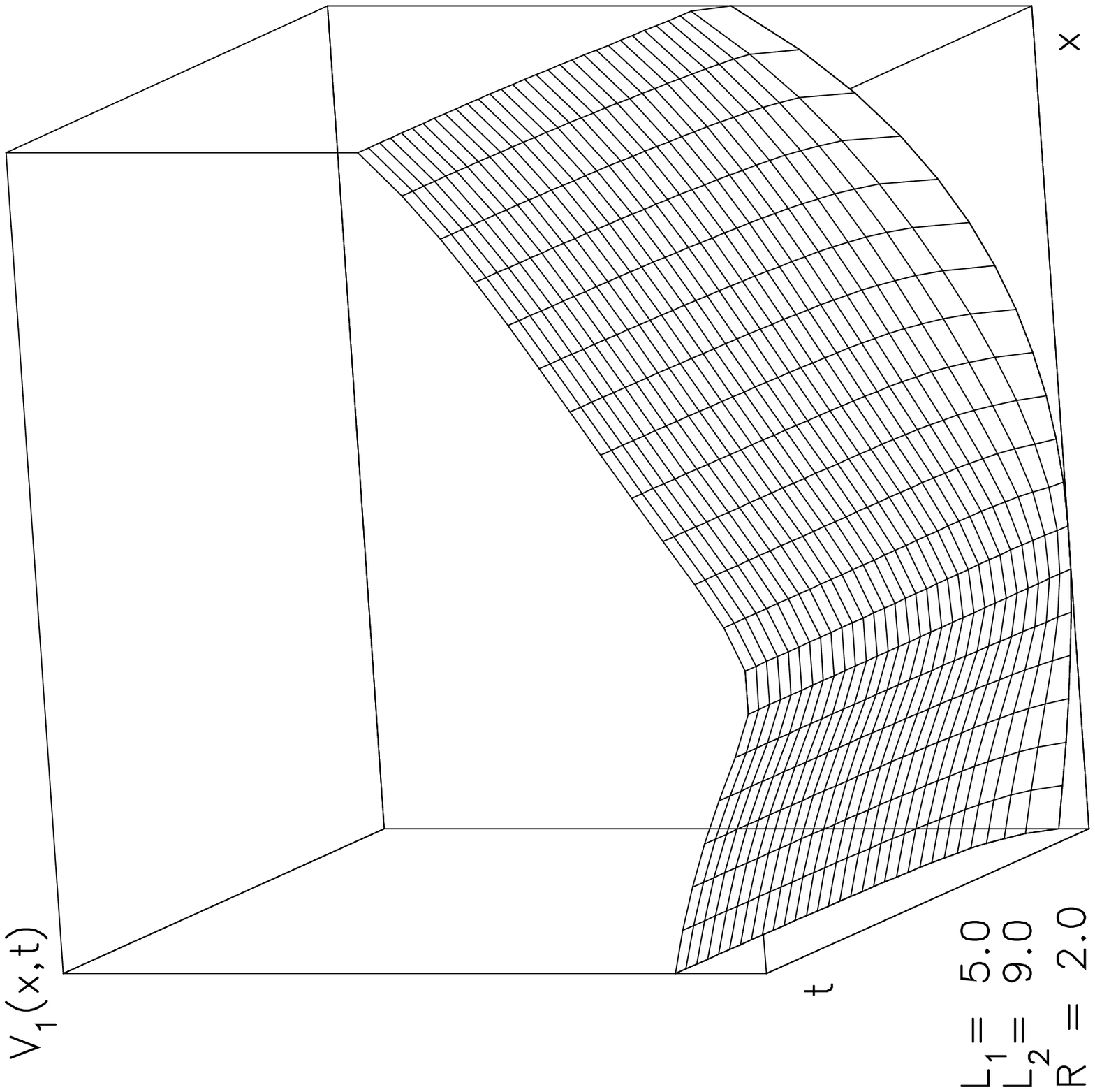}}}}
}\hfill
\parbox[b]{7.4cm}{
\epsfxsize=7.3cm 
\centerline{\rotate[r]{\hbox{\epsffile[28 28 570
556]{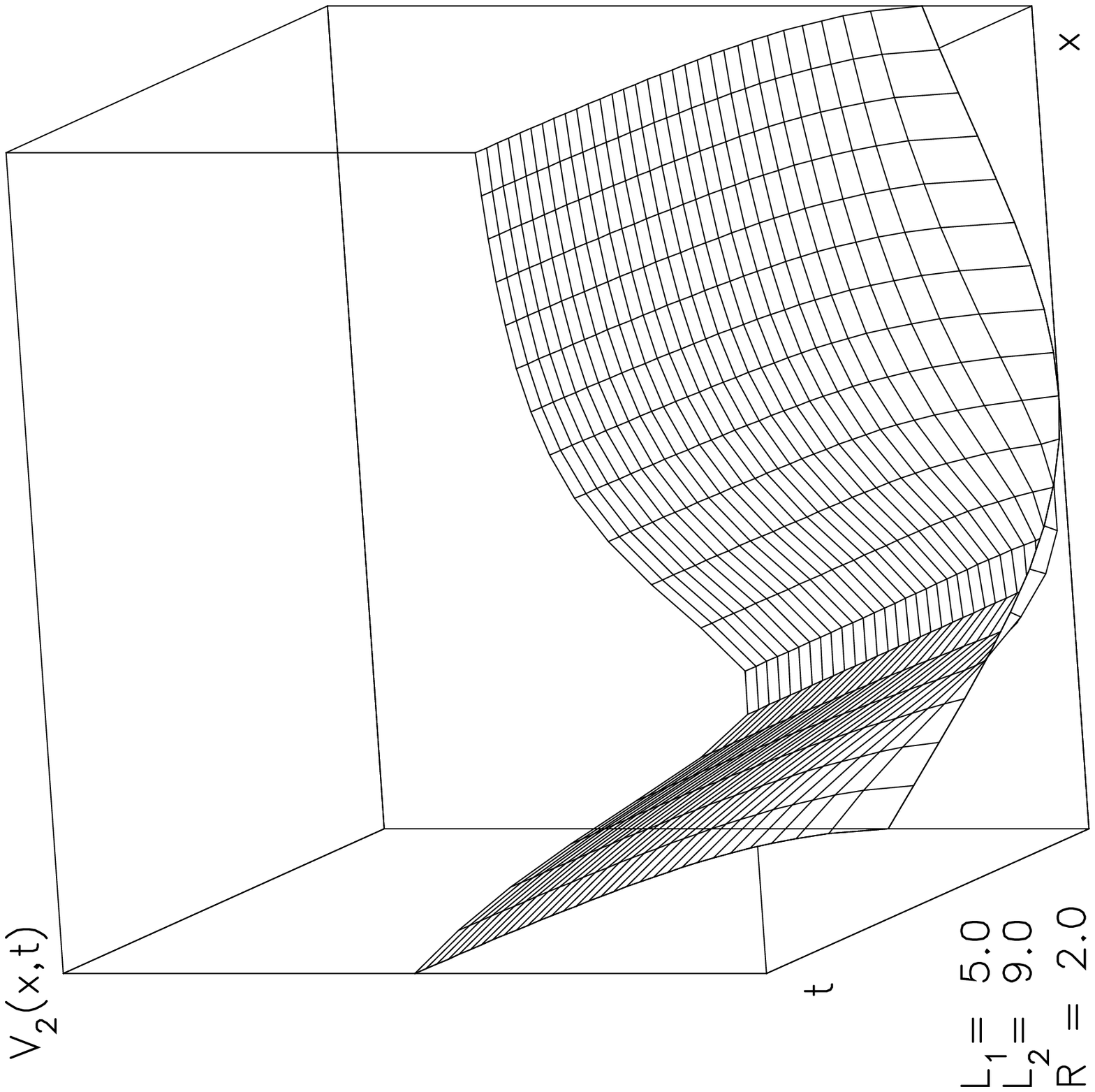}}}}
}
\parbox[b]{7.4cm}{
\epsfxsize=7.3cm 
\centerline{\rotate[r]{\hbox{\epsffile[28 28 570
556]{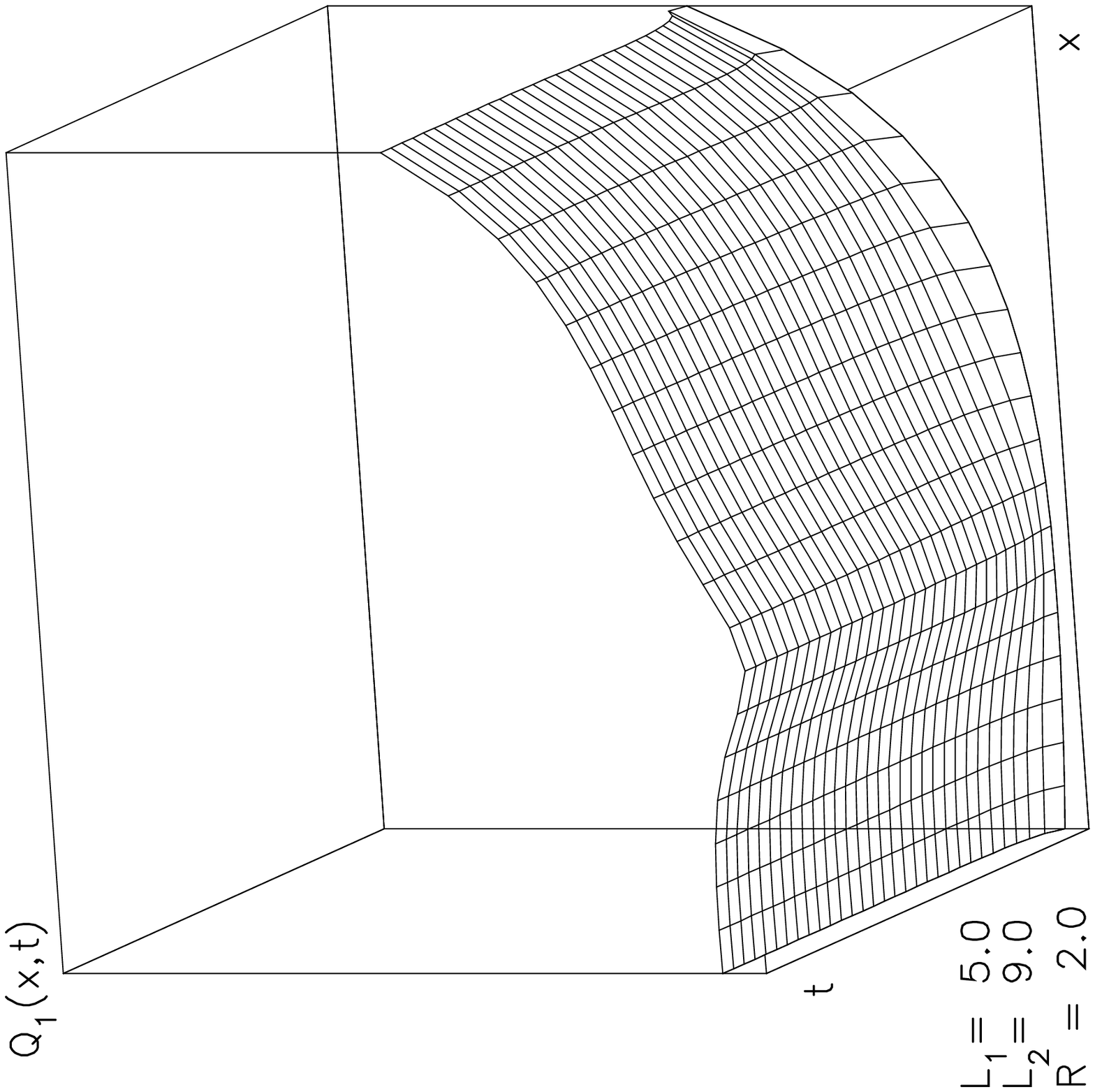}}}}
}\hfill
\parbox[b]{7.4cm}{
\epsfxsize=7.3cm 
\centerline{\rotate[r]{\hbox{\epsffile[28 28 570
556]{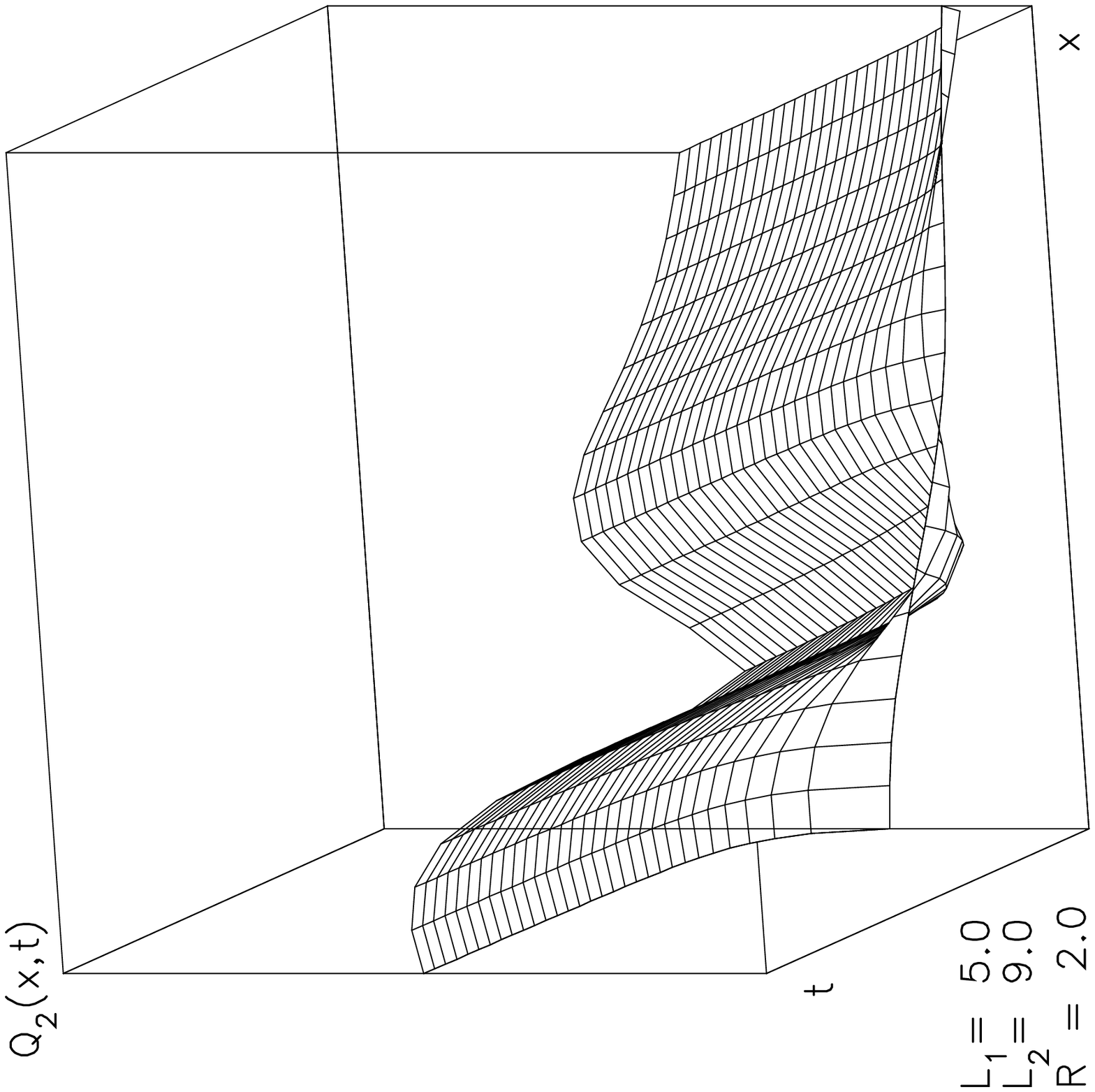}}}}
}
\parbox{15cm}{
\caption{As figure 3, but for high indifference $L_2$ in
subpopulation 2.\label{fi4}}
}
\end{figure}

\begin{figure}[htbp]
\parbox[b]{7.4cm}{
\epsfxsize=7.3cm 
\centerline{\rotate[r]{\hbox{\epsffile[28 28 570
556]{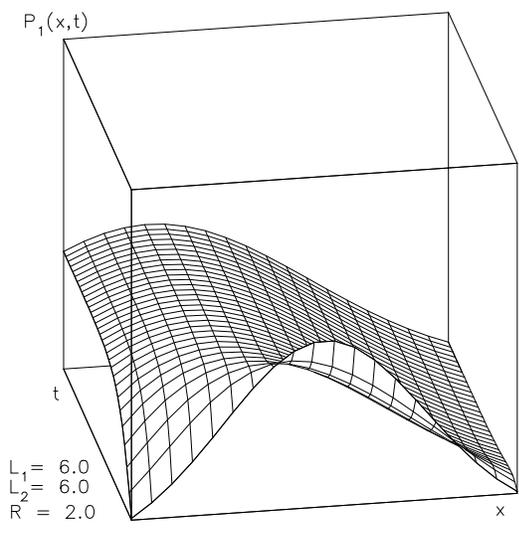}}}}
}\hfill
\parbox[b]{7.4cm}{
\epsfxsize=7.3cm 
\centerline{\rotate[r]{\hbox{\epsffile[28 28 570
556]{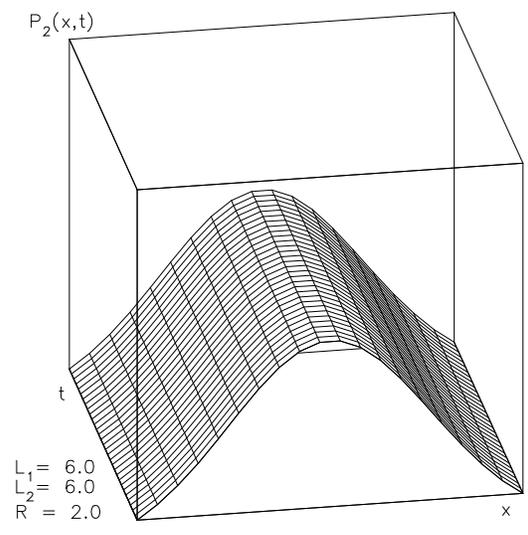}}}}
}
\parbox{15cm}{
\caption{Avoidance processes for one-sided dislike.\label{fi6}}
}
\end{figure}

In the following, the {\sc Boltzmann-Fokker-Planck} equations
for behavioral changes will be illustrated by representative computer
simulations. We shall examine the case of $A=2$ subpopulations,
and situations for which the interesting aspect of the individual behavior 
can be described
by a certain {\em position} $x \in [1/20,1]$ on a one-dimensional continuous
scale (i.e., $n=1$, $\vec{x} \equiv x$). Then, the integrability conditions
(\ref{intcond}) are automatically fulfilled, and the social field
\begin{equation}
 V_a(x,t) = - \int\limits_{x_0}^x dx' \, K_a(x',t) - c_a(t)
\end{equation}
is well-defined. The parameter $c_a(t)$ can be chosen arbitrarily. 
We will take for $c_a(t)$ the value that shifts the absolute
minimum of $V_a(x,t)$ to zero, that means,
\begin{equation}
 c_a(t) := \min_x \left( - \int\limits_{x_0}^x dx' \, K_a(x',t) \right) \, .
\end{equation}
\begin{itemize}
\item Since we will restrict the simulations to the case of
imitative or avoidance processes, the shape of the social field
$V_a(x,t)$ changes with time only due to changes of the probability
distributions $P_a(x,t)$ (cf. (\ref{SUM})), 
that means, due to behavioral changes
of the individuals (see figures \ref{fi1} to \ref{fi6}).
\end{itemize}
In the one-dimensional case one can find the formal {\em stationary solution}
\begin{equation}
 P_a(x) = P_a(x_0) \frac{Q_{a}(x_0)}{Q_{a}(x)}
\exp\left( 2 \int\limits_{x_0}^x dx' \, \frac{K_{a}(x')}{Q_{a}(x')}
\right) \, ,
\label{ABh}
\end{equation}
which we expect to be approached in the limit of large times
$t \rightarrow \infty$. Due to the dependence of $K_a(x)$ and $Q_a(x)$
on $P_a(x)$, equations (\ref{ABh}) are only {\em implicit} equations.
However, from (\ref{ABh}) we can derive the following conclusions:
\begin{itemize}
\item If the diffusion coefficients are constant ($Q_a(x) \equiv Q_a$),
(\ref{ABh}) simplifies to
\begin{equation}
 P_a(x) = P_a(x_0) \exp \left( - \frac{2}{Q_a} \Big[ V_a(x) + c_a \Big]
 \right) \, ,
\end{equation}
that means, the stationary solution $P_a(x)$ is completely determined
by the social field $V_a(x)$. Especially, $P_a(x)$ has its maxima at the
positions $x_a^*$, where the social field $V_a(x)$ has its minima
(see fig. \ref{fi1}).
The diffusion constant $Q_a$ regulates the width of the behavioral
distribution $P_a(x)$. 
\item If the diffusion coefficients $Q_a(x)$ are varying functions of the
position $x$, the behavioral distribution $P_a(x)$
is also influenced by the
concrete form of $Q_a(x)$. From (\ref{ABh}) one expects
high behavioral probabilities $P_a(x)$ where the diffusion coefficients
$Q_a(x)$ are small (see fig. \ref{fi2}, where the probability distribution
$P_1(x)$ cannot be explained solely by the social field $V_1(x)$).
\item Since the stationary solution $P_a(x)$ depends on both, $K_a(x)$ and
$Q_a(x)$, different combinations of $K_a(x)$ and $Q_a(x)$ can lead to the
same probability distribution $P_a(x)$ (see fig. \ref{fi4}
in the limit of large times).
\end{itemize}
For the following simulations, we shall use the {\em ansatz}\alpheqn{Ans} 
\begin{equation}
 R^a(x'|x;t) = \frac{\mbox{e}^{U^a(x',t) - U^a(x,t)}}
 {D_a(x',x;t)} 
\end{equation}
for the readiness $R^a(x'|x;t)$ to change from $x$ to $x'$
(cf. (\ref{util})).
With the utility function
\begin{equation}
 U^a(x,t) := - \frac{1}{2} \left( \frac{x - x_a}{l_a} \right)^2
 \, , \qquad l_a := \frac{L_a}{20}
\end{equation}
subpopulation $a$ prefers behavior $x_a$. $L_a$ means the
{\em indifference} of subpopulation $a$
with respect to variations of the position $x$.
Moreover, we take
\begin{equation}
 \frac{\nu_{ab}(t)}{D_a(x',x;t)} := \mbox{e}^{-|x' - x|/r} \, ,
 \qquad r = \frac{R}{20} \, ,
 \label{Dist}
\end{equation}\reseteqn
where $R$ can be interpreted as measure for the {\em range of interaction}.
According to (\ref{Dist}), the rate of behavioral changes is the smaller
the greater they are. Only small changes of the position (i.e., between
neighboring positions) contribute with an appreciable rate.

\subsection{Sympathy and interaction frequency} \label{s4.1}

Let $s_{ab}(t)$ be the degree of {\em sympathy} 
which individuals of subpopulation
$a$ feel towards individuals of subpopulation $b$. 
Then, one expects the following:
Whereas the frequency $f_{ab}^1(t)$ of imitative processes will be
increasing with $s_{ab}(t)$, the frequency $f_{ab}^2(t)$ of avoidance processes
will be decreasing with $s_{ab}(t)$.
This functional relationship can, for example, be described by
\begin{eqnarray}
 f_{ab}^1(t) &:=& f_a^1(t) \, s_{ab}(t) \, , \nonumber \\
 f_{ab}^2(t) &:=& f_a^2(t) \Big( 1 - s_{ab}(t) \Big)  
 \label{symp}
\end{eqnarray}
with
\begin{equation}
 0 \le s_{ab}(t) \le 1 \, .
\end{equation}
$f_a^1(t)$ is a measure for the frequency of imitative processes within
subpopulation $a$, $f_a^2(t)$ a measure for the frequency of avoidance
processes. If we assume the sympathy between individuals of the same
subpopulation to be be maximal, we have $s_{11}(t) \equiv 1 \equiv s_{22}(t)$.

\subsection{Imitative processes ($f_a^1(t) \equiv 1$, $f_a^2(t) \equiv 0$)}
\label{s4.2}
In the following simulations of imitative processes we assume the prefered
positions to be $x_1 = 6/20$ and $x_2 = 15/20$. With 
\begin{equation}
 \Big( s_{ab}(t) \Big) \equiv \Big( f_{ab}^1(t) \Big) := \left(
\begin{array}{cc}
1 & 1 \\
0 & 1 
\end{array} \right) \, ,
\end{equation}
the individuals of subpopulation $a=1$ like the individuals of subpopulation
$a=2$, but not the other way round. That means,
subpopulation 2 influences subpopulation 1, but not vice versa.
One could say, the individuals of subpopulation 2 act as {\em trendsetters}.
\par
As expected, in both behavioral distributions 
$P_a(x,t)$ there appears a maximum
around the prefered behavior $x_a$. In addition, due to imitative processes of
subpopulation 1, a second maximum of $P_1(x,t)$ develops around the
prefered behavior $x_2$ of the trendsetters. 
This second maximum is small, if the indifference $L_1$ of subpopulation 1
with respect to variations of the position $x$ is low (see fig. \ref{fi1}).
For high values of the indifference 
$L_1$ even the {\em majority} of individuals
of subpopulation 1 imitates the behavior of the trendsetters (see fig. 
\ref{fi2})!
\par
We shall now consider the case
\begin{equation}
 \Big( s_{ab}(t) \Big) \equiv \Big( f_{ab}^1(t) \Big) := \left(
\begin{array}{cc}
1 & 1 \\
1 & 1 
\end{array} \right) \, ,
\end{equation}
for which the subpopulations influence 
each other mutually with equal strengths.
If the indifference $L_a$ with respect to changes of the position $x$
is small in both subpopulations~$a$, {\em each} probability distribution 
$P_a(x,t)$ has {\em two} maxima. The higher maximum is located around the 
prefered position $x_a$. A second maximum can be found around the position
prefered in the {\em other} subpopulation. It is the higher, the greater
the indifference $L_a$ is (see fig. \ref{fi3}).
\par
However, if $L_a$ exceeds a certain value in at least one subpopulation,
only {\em one} maximum develops in each behavioral distribution $P_a(x,t)$!
Despite the fact,
that the social fields $V_a(x,t)$
and diffusion coefficients $Q_a(x,t)$ of the subpopulations $a$
are different because of their different prefered positions $x_a$ 
(and different utility functions $U^a(x,t)$), 
the behavioral distributions $P_a(x,t)$ agree after some time!     
Especially, the maxima $x_a^*$ of the distributions $P_a(x,t)$
are located at the {\em same} position $x^*$ in both subpopulations. 
$x^*$ is nearer to the position $x_a$ of the
subpopulation $a$ with the lower indifference $L_a$ (see fig. \ref{fi4}). 

\subsection{Avoidance processes ($f_a^1(t) \equiv 0$, $f_a^2(t) \equiv 1$)} 
\label{s4.3}
\begin{figure}[htbp]
\parbox[b]{7.4cm}{
\epsfxsize=7.3cm 
\centerline{\rotate[r]{\hbox{\epsffile[28 28 570
556]{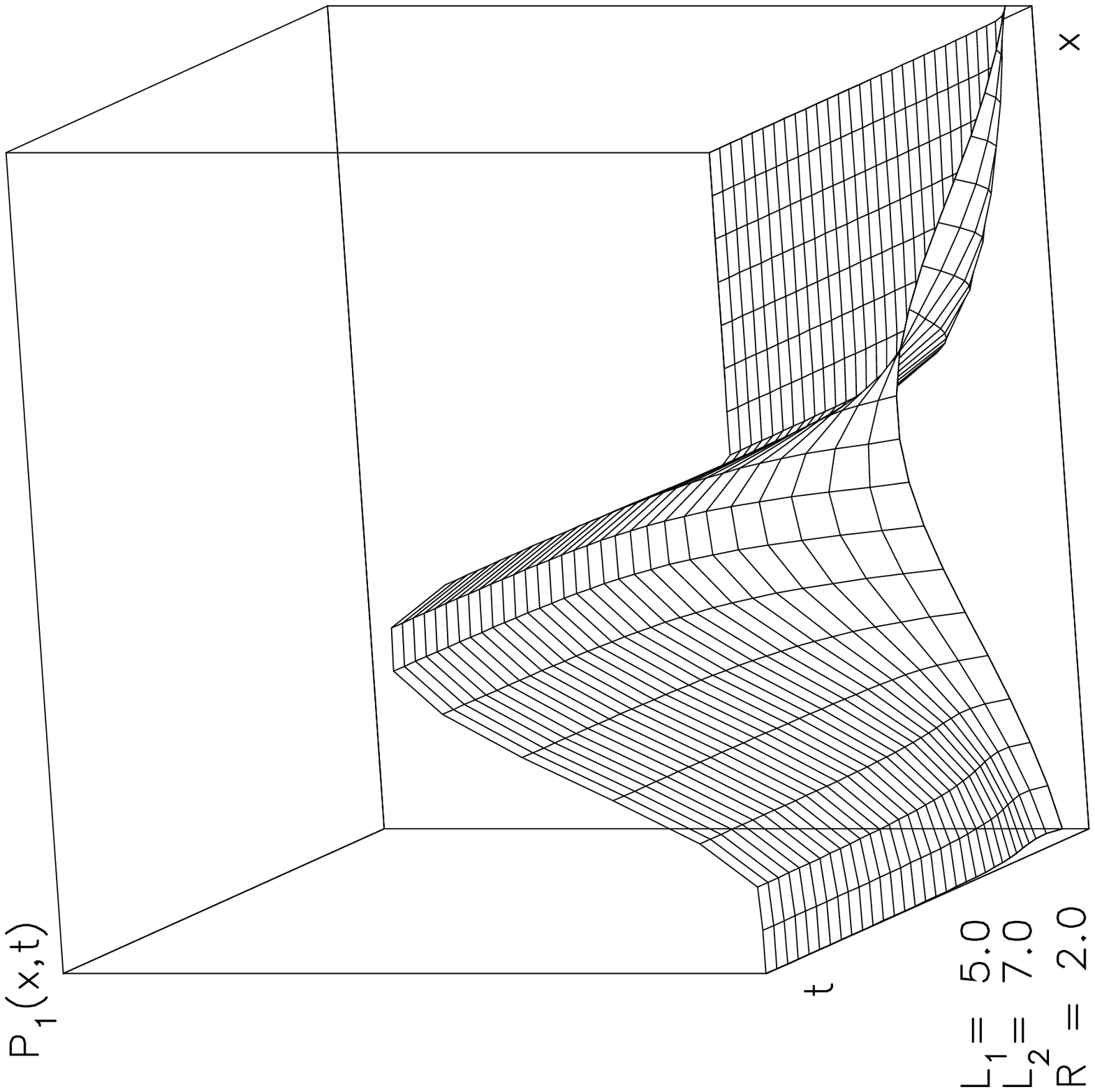}}}}
}\hfill
\parbox[b]{7.4cm}{
\epsfxsize=7.3cm 
\centerline{\rotate[r]{\hbox{\epsffile[28 28 570
556]{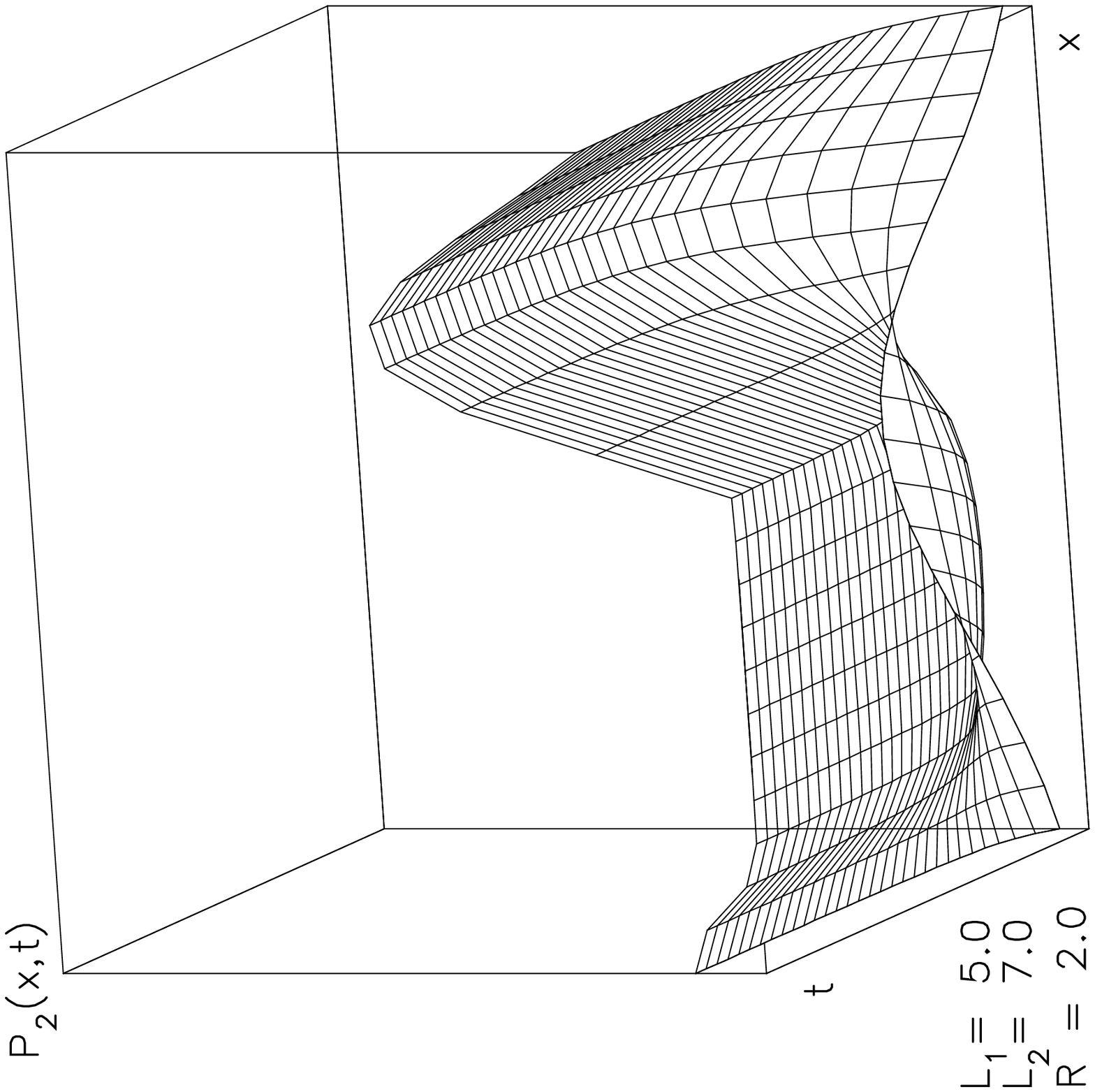}}}}
}
\parbox[b]{7.4cm}{
\epsfxsize=7.3cm 
\centerline{\rotate[r]{\hbox{\epsffile[28 28 570
556]{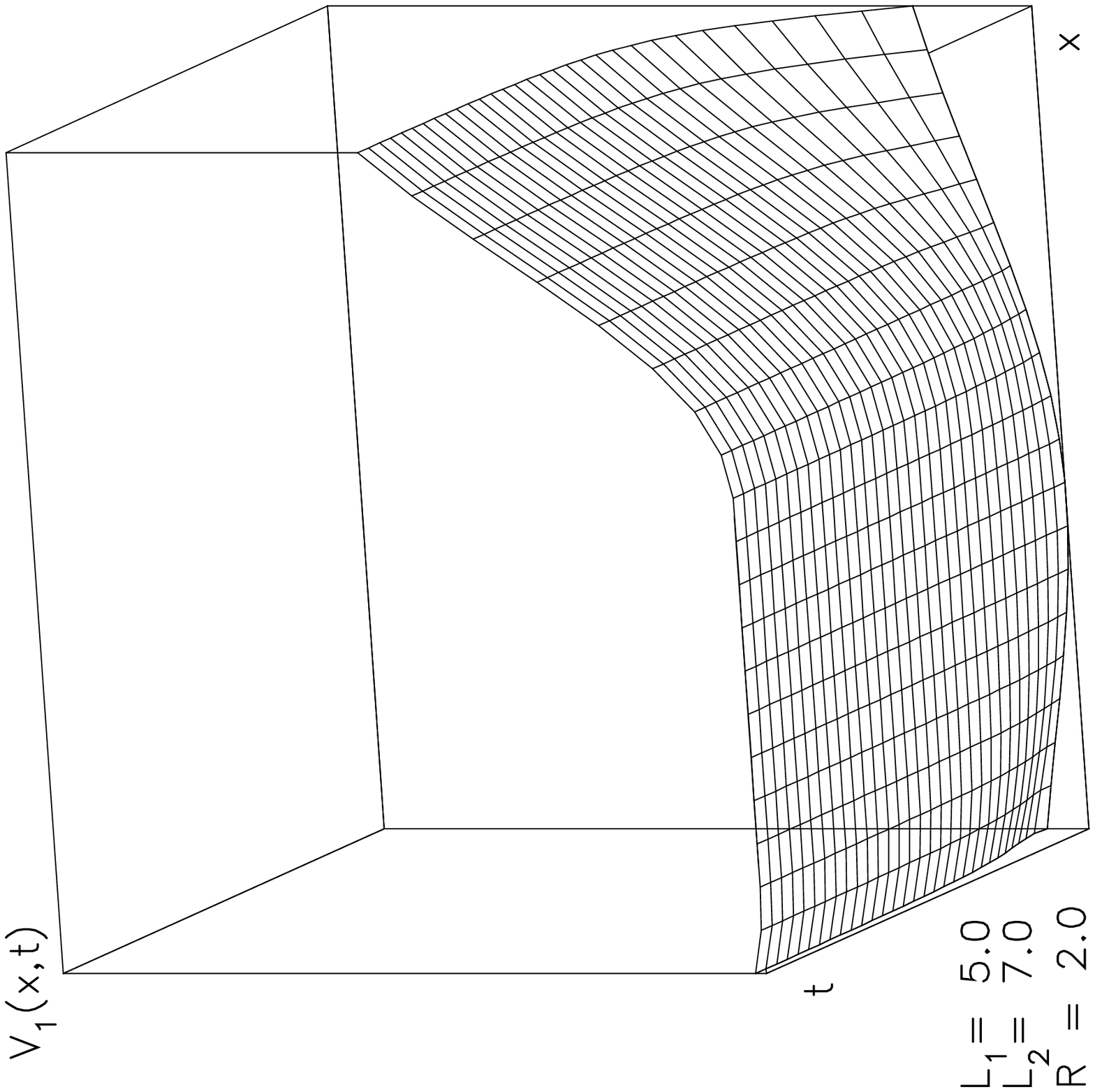}}}}
}\hfill
\parbox[b]{7.4cm}{
\epsfxsize=7.3cm 
\centerline{\rotate[r]{\hbox{\epsffile[28 28 570
556]{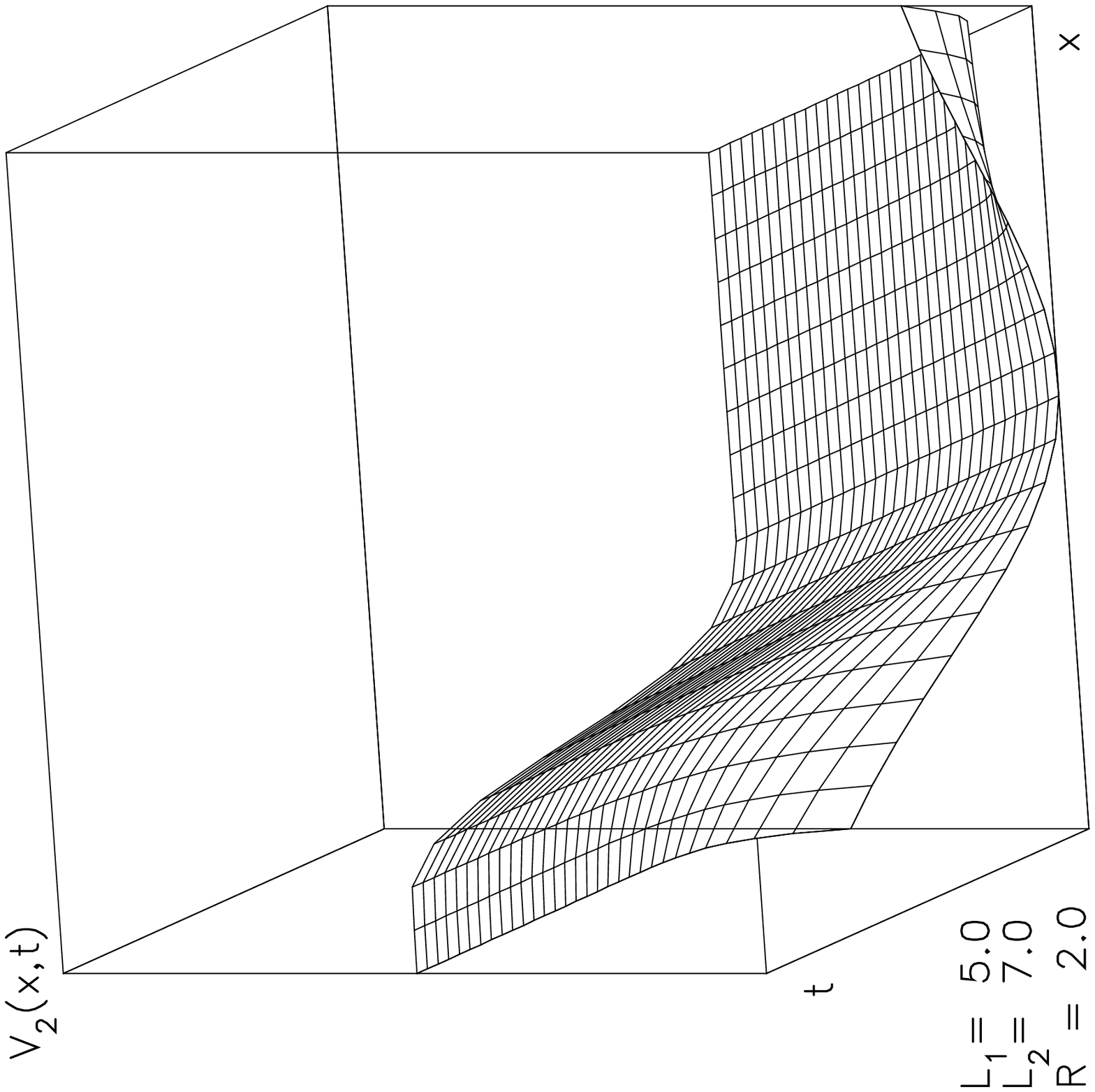}}}}
}
\parbox[b]{7.4cm}{
\epsfxsize=7.3cm 
\centerline{\rotate[r]{\hbox{\epsffile[28 28 570
556]{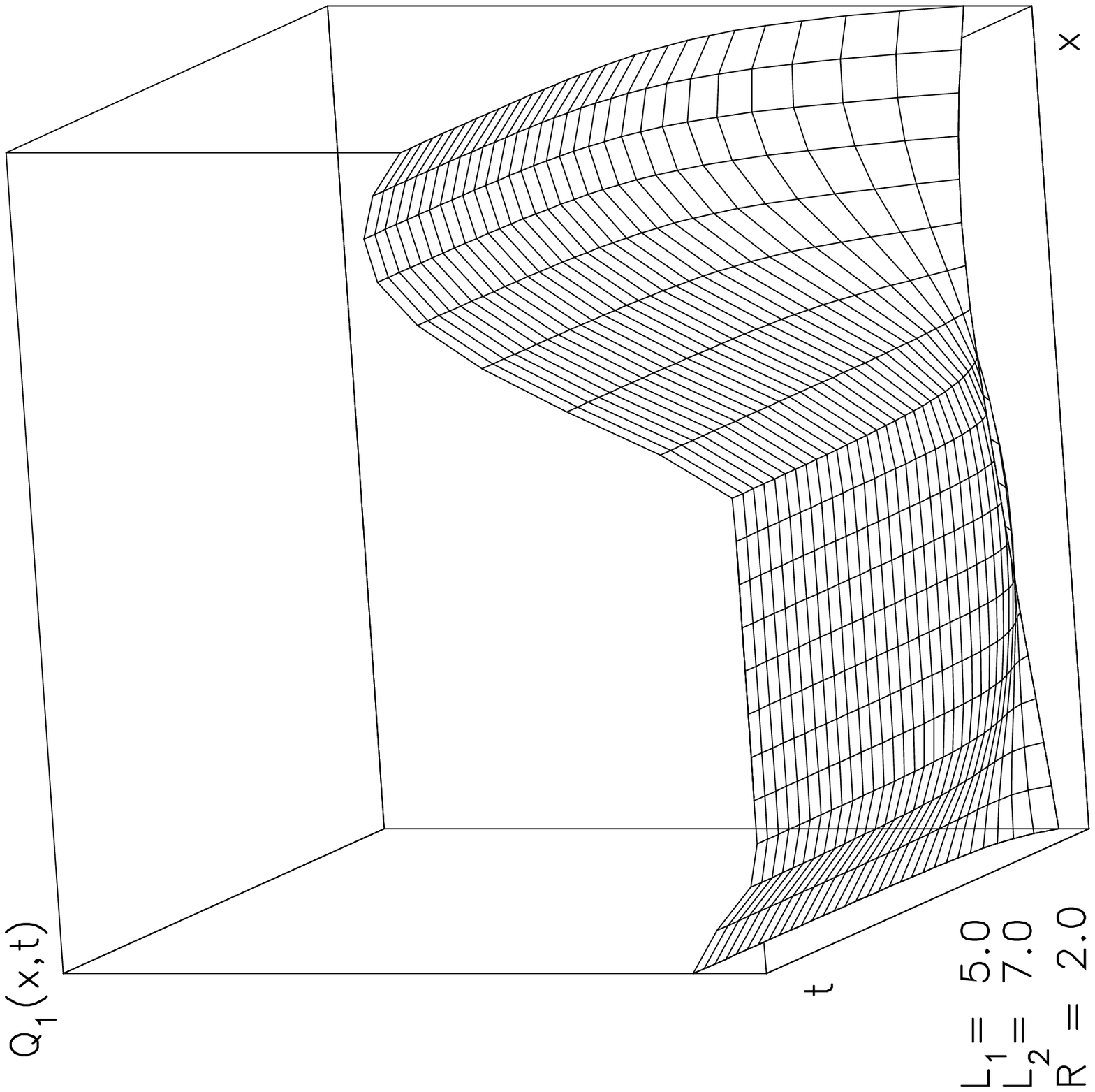}}}}
}\hfill
\parbox[b]{7.4cm}{
\epsfxsize=7.3cm 
\centerline{\rotate[r]{\hbox{\epsffile[28 28 570
556]{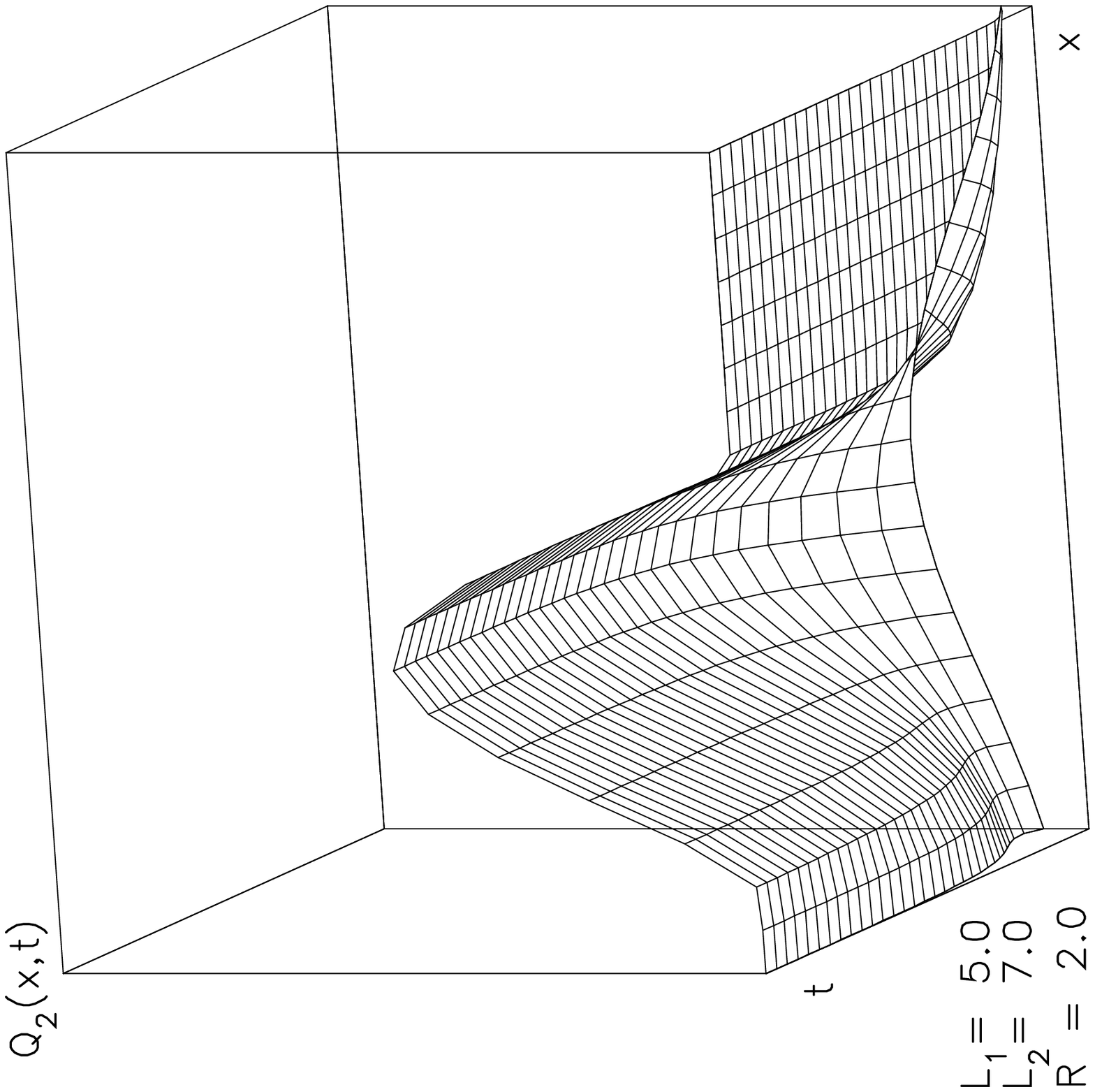}}}}
}
\parbox{15cm}{
\caption{Avoidance processes for mutual dislike of both
subpopulations.\label{fi5}}
}
\end{figure}

For the simulation of avoidance processes we assume with
$x_1 = 9/20$ and $x_2 = 12/20$ that both subpopulations nearly prefer the same
behavior. Figure \ref{fi5} shows the case, where the individuals of
different subpopulations dislike each other:
\begin{equation}
 \Big( s_{ab}(t) \Big) := \left(
\begin{array}{cc}
1 & 0 \\
0 & 1 
\end{array} \right) \, , \qquad \mbox{i.e.,} \qquad
 \Big( f_{ab}^2(t) \Big) \equiv \left(
\begin{array}{cc}
0 & 1 \\
1 & 0 
\end{array} \right) \, .
\end{equation}
This corresponds to a mutual influence of one 
subpopulation on the respective other. 
The computational results prove:
\begin{itemize}
\item The individuals avoid behaviors which can be found in the 
other subpopulation. 
\item The subpopulation $a=1$ with the lower indifference $L_1 < L_2$
is distributed around the prefered behavior $x_1$ and
pushes away the other subpopulation!
\end{itemize}
Despite the fact that the initial behavioral 
distribution $P_a(x,0)$ agrees in both subpopulations, there is nearly
no overlapping of $P_1(x,t)$ and $P_2(x,t)$ after some time. This
is a typical example for {\em polarization phenomena} in the society.
\par
In figure \ref{fi6}, we assume that the 
individuals of subpopulation 2 like the
individuals of subpopulation 1 and, therefore, do not react on the behaviors
in subpopulation 1 with avoidance processes:
\begin{equation}
 \Big( s_{ab}(t) \Big) := \left(
\begin{array}{cc}
1 & 0 \\
1 & 1 
\end{array} \right) \, , \qquad \mbox{i.e.,} \qquad
 \Big( f_{ab}^2(t) \Big) \equiv \left(
\begin{array}{cc}
0 & 1 \\
0 & 0 
\end{array} \right) \, .
\end{equation}
As a consequence, 
$P_2(x,t)$ remains unchanged with time, whereas $P_1(x,t)$
drifts away from the prefered behavior $x_1$ due to avoidance processes.
Surprisingly, the polarization effect 
is much smaller than in figure \ref{fi5}!
The distributions $P_1(x,t)$ and $P_2(x,t)$ overlap considerably. This is,
because the slope of $P_2(x,t)$ is smaller than in figure \ref{fi5}
(and remains constant). As a consequence, the probability for an individual
of subpopulation~1 to meet a disliked individual of subpopulation 2 with
the same behavior $x$ can hardly be decreased by a small behavioral change.
One may conclude, that polarization effects (which often lead to an
escalation) can be reduced, if individuals do not return dislike
by dislike.

\section{Empirical determination of the model parameters}\label{s5}

For practical purposes one has, of course, to determine the model parameters
from empirical data. Therefore, let us assume to know empirically the
distribution functions $P_a^{\rm e}(\vec{x},t_l)$, 
[the interaction rates $\nu_{ab}^{\rm e}(t_l)$] and the effective
transition rates $w_{\rm e}^a(\vec{x}'|\vec{x};t_l)$ ($\vec{x}' \ne \vec{x}$)
for a couple of times $t_l \in \{ t_0,\dots,t_L \}$. 
The corresponding effective
social fields $V_a^{\rm e}(\vec{x},t_l)$ and diffusion
coefficients $Q_{aij}^{\rm e}(\vec{x},t_l)$ are, then, 
easily obtained as\alpheqn{empi}
\begin{equation}
 V_a^{\rm e}(\vec{x},t_l) 
 := - \int\limits^{\weg{x}} d\vec{x}' \cdot
 \vec{K}_a^{\rm e}(\vec{x}',t_l) 
\end{equation}
with
\begin{equation}
K_{a i}^{\rm e}(\vec{x},t_l) 
:= \int\limits_\Omega \! d^n x' \, 
(x'_i - x_i) w_{\rm e}^a(\vec{x}'|\vec{x};t_l) \, ,
\end{equation}\reseteqn
and 
\begin{equation}
Q_{a i j}^{\rm e}(\vec{x},t_l) 
:= \displaystyle \int\limits_\Omega \! d^n x' \, (x'_i - x_i) (x'_j - x_j)
 w_{\rm e}^a(\vec{x}'|\vec{x};t_l) \, .
\end{equation}
Much more difficult is the determination of the utility functions
$U_a^{\rm e}(\vec{x},t_l)$, $U^a_{\rm e}(\vec{x},t_l)$, the distance functions
$D_a^{\rm e}(\vec{x}',\vec{x};t_l)$, and the rates $\nu_a^{\rm e}(t_l)$,
$\nu_{ab}^{\rm 1e}(t_l) := \nu_{ab}^{\rm e}(t_l) f_{ab}^{\rm 1e}(t_l)$, 
$\nu_{ab}^{\rm 2e}(t_l) := \nu_{ab}^{\rm e}(t_l) f_{ab}^{\rm 2e}(t_l)$. 
This can be done by numerical minimization of the {\em error function} 
\begin{equation}
 F := \sum_{a=1}^A \sum_{l=0}^L 
 \sum_{\weg{x}, \weg{x}'\in \Omega \atop (\weg{x}'\ne \weg{x})} 
 \frac{1}{2} \left\{ \left[
 w_{\rm e}^a(\vec{x}'|\vec{x};t_l) - \frac{1}{D_a(\vec{x}',\vec{x};t_l)}
 g_a(\vec{x}',\vec{x};t_l) \right] P_a^{\rm e}(\vec{x},t_l) \right\}^2 \, ,
 \label{error}
\end{equation}
for example with the method of {\em steepest descent} \cite{EnRe86}.    
In (\ref{error}), we have introduced the abbreviation
\begin{equation}
 g_a(\vec{x}',\vec{x};t_l) := \nu_a(t_l) \mbox{e}^{U_a(\weg{x}',t_l)
 - U_a(\weg{x},t_l)} 
 + \sum_{b=1}^A \Big[ \nu_{ab}^1(t_l) P_b^{\rm e}(\vec{x}',t_l)
 + \nu_{ab}^2(t_l) P_b^{\rm e} (\vec{x},t_l) \Big]
 \mbox{e}^{U^a(\weg{x}',t_l) - U^a(\weg{x},t_l)} \, .
\end{equation}
It turns out (cf. \cite{Diss}), that the rates $\nu_a(t_l)$ have to be taken
constant during the minimization process (e.g., $\nu_a(t_l) \equiv 1$),
whereas the parameters $U_a(\vec{x},t_l)$, $U^a(\vec{x},t_l)$, 
$\nu_{ab}^1(t_l):= \nu_{ab}^{\rm e}(t_l) f_{ab}^{\rm 1}(t_l)$ 
and $\nu_{ab}^2(t_l):= \nu_{ab}^{\rm e}(t_l) f_{ab}^{\rm 2}(t_l)$ 
are to be varied.
For $1/D_a(\vec{x}',\vec{x};t_l)$ one inserts\alpheqn{dista}
\begin{equation}
 \frac{1}{D_a(\vec{x}',\vec{x};t_l)} 
= \frac{n_a(\vec{x}',\vec{x};t_l)}{d_a(\vec{x}',\vec{x};t_l)}
\end{equation}
with
\begin{equation}
 n_a(\vec{x}',\vec{x};t_l) :=
w_{\rm e}^a(\vec{x}'|\vec{x};t_l) g_a(\vec{x}',\vec{x};t_l)
\Big[ P_a^{\rm e}(\vec{x},t_l) \Big]^2 \! + w_{\rm e}^a(\vec{x}|\vec{x}';t_l)
g_a(\vec{x},\vec{x}';t_l) \Big[ P_a^{\rm e}(\vec{x}',t_l) \Big]^2 
\end{equation}
and
\begin{equation}
 d_a(\vec{x}',\vec{x};t_l)
:= \Big[ g_a(\vec{x}',\vec{x};t_l) P_a^{\rm e}(\vec{x},t_l) \Big]^2 \!
+ \Big[ g_a(\vec{x},\vec{x}';t_l) P_a^{\rm e}(\vec{x}',t_l) \Big]^2 \, .
\end{equation}\reseteqn
(\ref{dista}) follows from the minimum 
condition for $D_a(\vec{x}',\vec{x};t_l)$
(cf. \cite{Diss}). 
\par
Since $F$ may have a couple of minima due to its nonlinearity, suitable
start parameters have to be taken. Especially, the numerically determined
rates $\nu_{ab}^1(t_l)$ and $\nu_{ab}^2(t_l)$ have to be
non-negative.
\par
If $F$ is minimal for the parameters 
$U_a(\vec{x},t_l)$, $U^a(\vec{x},t_l)$, 
$D_a(\vec{x}',\vec{x};t_l)$, $\nu_a(t_l)$,
$\nu_{ab}^1(t_l)$ and $\nu_{ab}^2(t_l)$, this is (as can easily be checked)
also true for the scaled parameters
\begin{eqnarray}
 U_a^{\rm e}(\vec{x},t_l) &:=& U_a(\vec{x},t_l) - C_a(t_l) \, , \nonumber \\
 U^a_{\rm e}(\vec{x},t_l) &:=& U^a(\vec{x},t_l) - C^a(t_l) \, , \nonumber \\
 D_a^{\rm e}(\vec{x}',\vec{x};t_l) &:=& \frac{D_a(\vec{x}',\vec{x};t_l)}
 {D_a(t_l)} \, , \nonumber \\
 \nu_a^{\rm e}(t_l) &:=& \frac{\nu_a(t_l)}{D_a(t_l)} \, , \nonumber \\
 \nu_{ab}^{1{\rm e}}(t_l) &:=& \frac{\nu_{ab}^1(t_l)}{D_a(t_l)} \, ,
 \nonumber \\
 \nu_{ab}^{2{\rm e}}(t_l) &:=& \frac{\nu_{ab}^2(t_l)}{D_a(t_l)} \, .
\end{eqnarray}
In order to obtain unique results we put
\begin{equation}
 \sum_{\weg{x}\in \Omega} U_a^{\rm e}(\vec{x},t_l) \stackrel{!}{\equiv} 0 \, ,
 \qquad \sum_{\weg{x}\in \Omega} U^a_{\rm e}(\vec{x},t_l) 
 \stackrel{!}{\equiv} 0 \, ,
\end{equation}
and
\begin{equation}
 \sum_{\weg{x}, \weg{x}'\in \Omega \atop (\weg{x}'\ne \weg{x})} 
 \frac{1}{D_a^{\rm e}(\vec{x}',\vec{x};t_l)}
 \stackrel{!}{\equiv} 
 \sum_{\weg{x}, \weg{x}'\in \Omega \atop (\weg{x}'\ne \weg{x})} 1 \, ,
\end{equation}
which leads to
\begin{equation}
 C_a(t_l) := \frac{\displaystyle \sum_{\weg{x}\in \Omega} U_a(\vec{x},t_l)}
 {\displaystyle \sum_{\weg{x}\in \Omega} 1} \, , \qquad
 C^a(t_l) := \frac{\displaystyle \sum_{\weg{x}\in \Omega} U^a(\vec{x},t_l)}
 {\displaystyle \sum_{\weg{x}\in \Omega} 1} \, ,
\end{equation}
and
\begin{equation}
 \frac{1}{D_a(t_l)} := \frac{\displaystyle 
 \sum_{\weg{x}, \weg{x}'\in \Omega \atop (\weg{x}'\ne\weg{x})} 
 \frac{1}{D_a(\vec{x}',\vec{x};t_l)}}
 {\displaystyle 
 \sum_{\weg{x}, \weg{x}'\in \Omega \atop (\weg{x}'\ne \weg{x})} 1} \, .
\end{equation}
$C_a(t_l)$ and $C^a(t_l)$ are {\em mean utilities}, whereas $D_a(t_l)$
is a kind of {\em unit of distance}.
\par
The distances $D_a^{\rm e}(\vec{x}',\vec{x};t)$ are suitable quantities for
{\em multidimensional scaling} \cite{Kr78,Yo87}. 
They reflect the ``psychical structure''
(psychical topology) of individuals of subpopulation $a$, 
since they determine which behaviors are
more or less related (compatible). 
By the dependence on 
$a$, $D_a^{\rm e}(\vec{x}',\vec{x};t)$ distinguishes 
different psychical structures
resulting in different types $a$ of behavior and, therefore, different
``characters''.

\section{Summary and outlook}\label{s6}

In this article, a behavioral model has been proposed that incorporates in a
consistent way many models of social theory:
the diffusion models, the multinomial logit model, {\sc Lewin}'s field
theory, the logistic equation, the gravity model, the
{\sc Weidlich-Haag} model, and the game dynamical equations.
This very general model opens new 
perspectives concerning a theoretical description and
understanding of behavioral changes, since it is formulated  fully 
mathematically. It takes into account spontaneous (or externally induced)
behavioral changes and behavioral changes due to pair interactions.
Two important kinds of pair interactions have been distinguished: imitative
processes and avoidance processes. 
The model turns out to be suitable for computational simulations,
but it can also be applied to concrete empirical data.

\subsection{Memory effects}\label{s6.1}

The formulation of the model in the previous sections has neglected {\em memory
effects} that may also influence behavioral changes. However, memory
effects can be easily included by generalizing the {\sc Boltzmann}-like
equations to\alpheqn{memb}
\begin{equation}
\frac{d}{dt}P_a(\vec{x},t) = 
\int\limits_{t_0}^t dt' \sum_{\weg{x}'\in \Omega} 
\Big[ w_{t-t'}^a(\vec{x}|\vec{x}';t')P_a(\vec{x}',t') 
- w_{t-t'}^a(\vec{x}'|\vec{x};t')P_a(\vec{x},t') \Big] 
\label{memboltz}
\end{equation}
with the effective transition rates
\begin{equation}
w_{t-t'}^a(\vec{x}'|\vec{x};t') := w^{t-t'}_a(\vec{x}'|\vec{x};t') 
+ \sum_{b=1}^A \sum_{\weg{y}\in \Omega} \sum_{\weg{y}'\in \Omega} 
w^{t-t'}_{ab}(\vec{x}',\vec{y}'|\vec{x},\vec{y};t')P_b(\vec{y},t') \, ,
\label{memrates}
\end{equation}\reseteqn
and generalizing the {\sc Boltzmann-Fokker-Planck} equations to\alpheqn{membfp}
\begin{equation}
\begin{array}{rcl}
\displaystyle \frac{\partial}{\partial t}P_a(\vec{x},t) 
&=& \displaystyle \int\limits_{t_0}^t dt' \, \bigg\{
- \sum_{i=1}^n \frac{\partial}{\partial x_{i}} \Big[K_{a i}^{t - t'}
(\vec{x},t') P_a(\vec{x},t')\Big] \\
& & \qquad \quad \displaystyle +  \frac{1}{2} \sum_{i, j=1}^n 
\frac{\partial}{\partial x_{i}}\frac{\partial}
{\partial x_{j}} \Big[Q_{a i j}^{t - t'}(\vec{x},t')
P_a(\vec{x},t')\Big] \bigg\}
\end{array} 
\label{memBFP}
\end{equation}
with the effective drift coefficients
\begin{equation}
K_{a i}^{t - t'}(\vec{x},t') 
:= \int\limits_\Omega \! d^n x' \, 
(x'_i - x_i) w^a_{t - t'}(\vec{x}'|\vec{x};t') \, ,
\label{memdrift}
\end{equation}
the effective diffusion coefficients
\begin{equation}
Q_{a i j}^{t - t'}(\vec{x},t') 
:= \int\limits_\Omega \! d^n x' \, (x'_i - x_i) (x'_j - x_j)
 w^a_{t - t'}(\vec{x}'|\vec{x};t') \, ,
\label{memdiff}
\end{equation}
and
\begin{equation}
w_{t-t'}^a(\vec{x}'|\vec{x};t') := w^{t-t'}_a(\vec{x}'|\vec{x};t') 
+ \sum_{b=1}^A \int\limits_\Omega d^n y \int\limits_\Omega d^n y' \, 
w^{t-t'}_{ab}(\vec{x}',\vec{y}'|\vec{x},\vec{y};t')P_b(\vec{y},t') \, .
\end{equation}\reseteqn
Obviously, in these formulas there only appears
an additional integration over past
times $t'$ \cite{Diss}. The influence of the past results in a dependence
of $w_{t-t'}^a(\vec{x}'|\vec{x};t')$, $K_{a i}^{t - t'}(\vec{x},t')$, 
and $Q_{a i j}^{t - t'}(\vec{x},t')$ on $(t-t')$. The 
{\sc Boltzmann}-like equations (\ref{Boltzlike}) resp. the 
{\sc Boltzmann-Fokker-Planck} equations (\ref{bfp}) used in the
previous sections result from (\ref{memb}) resp.
(\ref{membfp}) in the {\sc Markov}ian limit 
\begin{equation}
 w_{t-t'}^a(\vec{x}'|\vec{x};t') := w^a(\vec{x}'|\vec{x};t)
 \delta (t - t')
\end{equation}
of short memory (where $\delta (.)$ is the {\sc Dirac} delta function).

\subsection{Analogies with chemical reactions}

The {\sc Boltzmann}-like equations (\ref{Boltzlike}) can also be used
for the description of chemical reactions, where the states $\vec{x}$
denote the different sorts of molecules (or atoms), and
$a$ distinguishes different isotopes or conformeres. Imitative and
avoidance processes correspond in chemistry
to self-activatory and self-inhibitory reactions. Although the concrete
transition rates will be different from (\ref{concrates}), (\ref{util}) 
in detail, there may be found analogous results for chemical reactions.
Note, that the {\sc Arrhenius} formula for the rate of chemical reactions
\cite{Ga85} can be put into a form similar to
(\ref{util}) \cite{Diss}.
\clearpage
\paragraph{Acknowledgements}\mbox{ }\\[7mm]
This work has been financially supported by the {\em Volkswagen Stiftung}
and the {\em Deutsche Forschungsgemeinschaft} (SFB 230).
The author is grateful to Prof. Dr. W. Weidlich and Dr. R. Reiner for
valuable discussions and commenting on the manuscript.

%
%
\clearpage
\thispagestyle{empty}
\clearpage
\thispagestyle{empty}
\clearpage
\thispagestyle{empty}
\clearpage
\thispagestyle{empty}
\clearpage
\thispagestyle{empty}
\clearpage
\thispagestyle{empty}
\end{document}